\documentclass[letterpaper,twocolumn,10pt]{article}
\usepackage{usenix2019,epsfig,endnotes}

\usepackage{listings} 
\usepackage{multirow}
\usepackage{graphicx}
\usepackage{comment}
\usepackage[frozencache,cachedir=.]{minted}
\usepackage{xcolor}
\usepackage{array}
\usepackage{amssymb}% http://ctan.org/pkg/amssymb
\usepackage{pifont}% http://ctan.org/pkg/pifont
\newcolumntype{R}[1]{>{\begin{minipage}{#1}\raggedleft\let\newline\\
\arraybackslash}m{#1}<{\end{minipage}}}
\usepackage{tabularx,booktabs} % For formal tables

\providecommand{\keywords}[1]{\textbf{\textit{Index terms---}} #1}

\usepackage{xspace}

\newcounter{obstacle}[obstacle]

\usepackage{url}
\usepackage{booktabs, multirow, array, environ}
\usepackage[ruled,linesnumbered]{algorithm2e} % For algorithms

\SetAlFnt{\small}
\SetAlCapFnt{\small}
\SetAlCapNameFnt{\small}
\SetAlCapHSkip{0pt}
\IncMargin{-\parindent}

\usepackage{color}

\definecolor{codegreen}{rgb}{0,0.6,0}
\definecolor{codegray}{rgb}{0.5,0.5,0.5}
\definecolor{codepurple}{rgb}{0.58,0,0.82}
\definecolor{backcolour}{rgb}{0.95,0.95,0.92}

\lstdefinestyle{mystyle}{
  commentstyle=\color{codegreen},
  keywordstyle=\color{magenta},
  numberstyle=\tiny\color{codegray},
  stringstyle=\color{codepurple},
  basicstyle=\footnotesize,
  breakatwhitespace=false,         
  breaklines=true,                 
  captionpos=b,                    
  keepspaces=true,                 
  numbers=left,                    
  numbersep=5pt,                  
  showspaces=false,                
  showstringspaces=false,
  showtabs=false,                  
  tabsize=2
}

\newcommand{\EnFuzz}{\texttt{EnFuzz}}

\newcommand{\toolname}{\texttt{EnFuzz} }
\newcommand{\toolOne}{\texttt{EnFuzz-A}}
\newcommand{\toolTwo}{\texttt{EnFuzz-L}}
\newcommand{\toolThree}{\texttt{EnFuzz}}
\newcommand{\toolFour}{\texttt{EnFuzz$^{-}$}}
\newcommand{\toolFive}{\texttt{EnFuzz-Q}}
\newcommand{\cvenum}{44 }
\newcommand{\bugnum}{60 }

\begin{document}

%don't want date printed
\date{}

%make title bold and 14 pt font (Latex default is non-bold, 16 pt)
\title{\Large \bf EnFuzz: Ensemble Fuzzing with Seed Synchronization among Diverse Fuzzers}

%for single author (just remove % characters)
\author{
{\rm Yuanliang Chen, Yu Jiang* \{jiangyu198964@126.com\},}\\
{\rm Fuchen Ma, Jie Liang, Mingzhe Wang, Chijin Zhou and hZuo Su}\\
School of Software, Tsinghua University, KLISS, Beijing, China
\and
{\rm Xun Jiao \{canbyjiaoxun@163.com\}}\\
ECE department of Villanova University
% copy the following lines to add more authors
% \and
% {\rm Name}\\
%Name Institution
} % end author

%\title{EnFuzz: Improving Generalization of Fuzzers Through Ensemble Approach}
%\title{EnFuzz: From Ensemble Learning to Ensemble Fuzzing
\maketitle

% Use the following at camera-ready time to suppress page numbers.
% Comment it out when you first submit the paper for review.
\thispagestyle{empty}

\subsection*{Abstract}
Fuzzing is widely used for vulnerability detection.
There are various kinds of fuzzers with different fuzzing strategies, and most of them perform well on their targets. 
However, in industrial practice, it is found that the performance of those well-designed fuzzing strategies is challenged by the complexity and diversity of real-world applications. 
In this paper, 
we systematically study an ensemble fuzzing approach.
%inspired by the idea of ensemble learning, we propose an ensemble fuzzing approach which we refer as \EnFuzz.
First, we define the diversity of base fuzzers in three heuristics: diversity of coverage information granularity, diversity of input generation strategy and diversity of seed selection and mutation strategy. Based on those heuristics, we choose several of the most recent base fuzzers that are as diverse as possible, and propose a globally asynchronous and locally synchronous (GALS) based seed synchronization mechanism to seamlessly ensemble those base fuzzers and obtain better performance. 
For evaluation, we implement \EnFuzz ~based on several widely used fuzzers such as QSYM and FairFuzz, and then test them on LAVA-M and Google's fuzzing-test-suite, which consists of 24 widely used real-world applications. This experiment indicates that, under the same constraints for resources, these base fuzzers perform differently on different applications, while EnFuzz always outperforms others in terms of path coverage, branch coverage and bug discovery. Furthermore, \EnFuzz ~found \bugnum new vulnerabilities in several well-fuzzed projects such as libpng and libjpeg, and \cvenum new CVEs were assigned.
%For evaluation, we implement \EnFuzz ~based on several widely used fuzzers, 
%(including AFL, AFLFast, FairFuzz, libFuzzer, Radamsa and QSYM),
%and then test them on two benchmarks and several real projects under the same resources constraint. Specially, on Google's fuzzing-test-suite consisting of widely used real-world application with code base 80K-220K LOCs, the experiment indicates that, these base fuzzers perform variously on different applications, while \EnFuzz ~always outperforms others in terms of path coverage, branch coverage and bug discovery.
%Compared with AFL, AFLFast, FairFuzz, QSYM, libFuzzer and Radamsa. \EnFuzz ~discovers 76.4\%, 140\%, 100\%, 81.8\%, 66.7\% and 93.5\% more bugs, executes 42.4\%, 61.2\%, 45.8\%, 66.4\%, 29.5\% and 44.2\% more paths and covers 15.5\%, 17.8\%, 12.9\%, 26.1\%, 19.9\% and 14.8\% more branches respectively. 
%For the result on LAVA-M with code base 2K-4K LOCs, it outperforms each base fuzzers as well.
%Furthermore, \EnFuzz ~found \bugnum new vulnerabilities in several  well-fuzzed projects such as libpng and libjpeg, while other base fuzzers only detect 35 new vulnerabilities at most, and \cvenum new CVEs were assigned.
%for image processing, and libiec61850 for device communication
%, as presented in the appendix.

\keywords{Ensemble Fuzzing, Seed Synchronization}

\section{Introduction}
  %Vulnerabilities of memory security are increasing risks in software, which hackers can exploit to cause many severe threats to programs such as confidential information leakage. 
Fuzzing is one of the most popular software testing techniques for bug and vulnerability detection. There are many fuzzers for academic and industrial usage. The key idea of fuzzing is to generate plenty of inputs to execute the target application and monitor for any anomalies.
%As the kernel of a fuzzer, 
While each fuzzer develops its own specific fuzzing strategy to generate inputs,
there are in general two main types of strategies. One is a generation-based strategy which uses the specification of input format, e.g. grammar, to generate complex inputs. For example, IFuzzer \cite{ifuzzer} takes a context-free grammar as specification to generate parse trees for code fragments.
Radamsa \cite{helin2016radamsa} reads sample files of valid data and generates interesting different outputs from them.
The other main strategy is a mutation-based strategy. This approach generates new inputs by mutating the existing seeds (good inputs contributing to improving the coverage).
Recently, mutation-based fuzzers are proposed to use coverage information of target programs to further improve effectiveness for bug detection. For example, libFuzzer \cite{libFuzzer} mutates seeds by utilizing the SanitizerCoverage \cite{SanitizerCoverage} instrumentation to track block coverage, while AFL \cite{afl} mutates seeds by using static instrumentation to track edge coverage. 

Based on the above mentioned two fuzzers, researchers have performed many optimizations. For example, AFLFast \cite{aflfast} improves the fuzzing strategy of AFL by selecting seeds that exercise low-frequency paths for additional mutations, and FairFuzz \cite{FairFuzz} optimizes AFL's mutation algorithm to prioritize seeds that hit rare branches. AFLGo \cite{aflgo} assigns more mutation times to the seeds closer to target locations. QSYM \cite{qsym} uses a practical concolic execution engine to solve complex branches of AFL.
All of these optimized fuzzers outperform AFL on their target applications and have already detected a large number of software bugs and security vulnerabilities. 

However, when we apply these optimized fuzzers to some real-world applications, these fuzzing strategies are inconsistent in their performance, their effectiveness on different applications varies accordingly. For example, in our evaluation on 24 real-world applications, AFLFast and FairFuzz perform better than AFL on 19 applications, while AFL performs better on the other 5 applications. Compared with AFL, libFuzzer performs better on 17 applications but worse on the other 7 applications.
%For parallel mode which is widely used in industry, FairFuzz and AFLFast only execute 80\% and 92\% paths, cover 95\% and 94\% branches of AFL. 
For the parallel mode of fuzzing which is widely-used in industry, AFLFast and FairFuzz  only detected  73.5\%  and 88.2\% of the unique bugs of AFL. 
These results show that the performance of existing fuzzers is challenged by the complexity and diversity of real-world applications. For a given real-world application, we cannot evaluate which fuzzer is better unless we spend significant time analyzing them or running each of these fuzzers one by one. This would waste a lot of human and computing resources \cite{klees2018evaluating}. 
This indicates that many of the current fuzzing strategies have a lack of robustness --- the property of being strong and stable consistently in constitution.  For industrial practice, more robust fuzzing strategies are desired when applied across different applications. 
%When it is applied into different applications, the term “robustness” refers to the consistent performance of fuzzers.
%Therefore, we need a fuzzer with the ability to provide more robust fuzzing across different applications.

\begin{comment}
The theory of ensemble learning \cite{zhou2015ensemble} proves that the generalization ability of an ensemble learner is usually much stronger than that of base learners \cite{hansen1990neural, schapire1990strength}. Similarly, we propose the idea of ensemble fuzzing, 
which intuitively helps in the following two aspects: \textit{robustness}, i.e. the consistent advantage on any application in the evaluation setup; \textit{performance}, i.e. achieving better metrics than any other fuzzers under the same resources constraint.

To demonstrate the effectiveness of an ensemble fuzzing, we need to deal with two main challenges before implementation:

\begin{itemize}
   \item [1.] \emph{Base Fuzzer Selection.} Base fuzzers are the underlying basics, and the diversity of base fuzzers is crucial to an ensemble fuzzer. The more diversity of these base fuzzers, the better the ensemble fuzzer performs. 
   \item [2.] \emph{Ensemble Architecture Design.} The architecture determines the effectiveness of ensemble fuzzing. The concrete ensemble architecture should be well designed, because a ensemble method should effectively combines these existing base fuzzers into a stronger ensemble fuzzer. 
\end{itemize}

To our best knowledge, we are the first to propose an ensemble fuzzing approach which we refer as \EnFuzz. 

\end{comment}

In this paper, we systematically study the performance of an ensemble fuzzing approach.
%inspired by the theory of ensemble learning which proves that the performance of an ensemble learner is usually much better than that of base learners \cite{zhou2015ensemble,hansen1990neural, schapire1990strength} and the global asynchronous and local synchronous(GALS) system design \cite{muttersbach2000practical}, we propose ensemble fuzzing.
First, we define the diversity of base fuzzers focusing on three heuristics: diversity of coverage information granularity, diversity of input generation strategy, as well as diversity of seed mutation and selection strategy. 
Then, we implement an ensemble architecture with a global asynchronous and local synchronous(GALS) based seed synchronization mechanism to integrate these base fuzzers effectively.
To enhance cooperation among different base fuzzers, the mechanism synchronizes interesting seeds(i.e., test cases covering new paths or triggering new crashes) periodically to all fuzzers running on the same target application. At the same time, it maintains a global coverage map to help collect those interesting seeds asynchronously from each base fuzzer. %In the meantime, i
%The second is the result integration mechanism, which completes the fuzzing session by collecting,de-duplicating and triaging results from all base fuzzers. 

For evaluation, we implement a prototype of \EnFuzz , based on several high-performance base fuzzers, including AFL, AFLFast, FairFuzz, QSYM, libFuzzer and Radamsa.
All fuzzers are repeatedly tested on two widely used benchmarks --- LAVA-M and Google's fuzzer-test-suite, following the kernel rules of evaluating fuzzing guideline\cite{klees2018evaluating}. %We follow the guidelines \cite{klees2018evaluating} for evaluation, except for the unique crash. We keep it for evaluation because although it is not the same as the unique bug, but demonstrates the path and probability to detect a bug. 
The average number of paths executed, branches covered and unique crashes discovered are used as metrics. The results demonstrate that, with the same resource usage, the base fuzzers perform differently on different applications, 
while \EnFuzz ~consistently and effectively improves the fuzzing performance. 
%For example, FairFuzz and AFLFast are evaluated to be fuzzers that always perform better than AFL in their test cases \cite{FairFuzz,aflfast}, but there are many cases that AFL performs better than those two fuzzers on those real-world applications. 
%The generalization limitations certainly exist in these base fuzzers, 
For example, there are many cases where the original AFL performs better on some real-world applications than the two optimized fuzzers FairFuzz and AFLFast. 
In all cases, the ensemble fuzzing always outperforms all other base fuzzers. 
%Even with little diversity, the ensemble approach achieves huge improvements, and the more diversity among these base fuzzers, the better ensemble fuzzing performs.

Specifically, on Google's fuzzer-test-suite consisting of real-world applications with a code base of 80K-220K LOCs, compared with AFL, AFLFast, FairFuzz, QSYM, libFuzzer and Radamsa, \EnFuzz ~discovers 76.4\%, 140\%, 100\%, 81.8\%, 66.7\% and 93.5\% more unique bugs, executes 42.4\%, 61.2\%, 45.8\%, 66.4\%, 29.5\% and 44.2\% more paths and covers 15.5\%, 17.8\%, 12.9\%, 26.1\%, 19.9\% and 14.8\% more branches respectively.
For the result on LAVA-M consisting of applications with a code base of 2K-4K LOCs, it outperforms each base fuzzer as well. For further evaluation on more widely used and several well-fuzzed open-source projects such as Libpng and jpeg, \EnFuzz ~finds \bugnum new real vulnerabilities, \cvenum of which are security-critical vulnerabilities and registered as new CVEs. However, other base fuzzers only detect 35 new vulnerabilities at most. %in the US National Vulnerability Database.

This paper makes the following main contributions:
\vspace{0.03in}
\begin{itemize}

	 \item [1.] While many earlier works have mentioned the possibility of using ensemble fuzzing, we are among the first to systematically investigate the practical ensemble fuzzing strategies and the effectiveness of ensemble fuzzing of various fuzzers. We evaluate the performance of typical fuzzers through a detailed empirical study. %following the kernel rules of the evaluating fuzzing guideline \cite{klees2018evaluating}. %We, thus, identify several valuable results, especially for their performance variation 
	 %We found that no base fuzzer consistently perform better than another on most real diverse applications, even as in most previous literature studies, one optimized fuzzer is usually evaluated to perform better than another on almost all applications or benchmarks such as LAVA-M.
	 We define the diversity of base fuzzers and study the effects of diversity on their performance. %of these base fuzzers.
% on large set of real-world applications

	 \item [2.] We implement a concrete ensemble approach with seed synchronization to improve the performance of existing fuzzers. \EnFuzz ~shows a more robust fuzzing practice across diverse real world applications. The prototype\footnote{\url{https://github.com/enfuzz/enfuzz}\label{footnote1}} is also scalable and open-source so as to integrate other fuzzers. %Furthremore, we implement an \EnFuzz ~server with a website interface\footnote{\url{http://wingtecher.com/Enfuzz/} ~~for free usage.} for free use.  

	 %\item [3.] We evaluate \EnFuzz ~on a widely used third-party benchmark consisting of real-world applications. The empirical results reveal the performance variation of existing fuzzers on different applications and demonstrate that the variation can be alleviated effectively by our ensemble approach. %It also demonstrates that the more diversity of base fuzzers, the better ensemble fuzzing performs.
	 
	 \item [3.] We apply \EnFuzz ~to fuzz several well-fuzzed projects such as libpng and libjpeg from GitHub, and several commercial products such as \texttt{libiec61850} from Cisco. Within 24 hours, \bugnum new security vulnerabilities were found and \cvenum new CVEs were assigned, while other base fuzzers only detected 35 new vulnerabilities at most. \EnFuzz~ has already been deployed in industrial practice, and more new CVEs are being reported$^{1}$.
 %So far, many people have signed up for trials, many new vulnerabilities have been discovered, and more CVEs are being applied for. 
\end{itemize}

\vspace{0.03in}
The rest of this paper is organized as follows:
Section \ref{Related Work} introduces related work. 
Section \ref{Motivating Example} illustrates ensemble fuzzing by a simple example.
Section \ref{Ensemble Fuzzing Approach EnFuzz} elaborates ensemble fuzzing, including the base fuzzer selection and ensemble architecture design.
Section \ref{Evaluation} presents the implementation and evaluation of \EnFuzz.
Section \ref{Discussion} discusses the potential threats of \EnFuzz, and we conclude in section \ref{Conclusion}. The appendix shows some empirical evaluations and observations. %about parallel fuzzing.

\section{Related Work}
\label{Related Work}
  Here below, we introduce the work related to generation-based fuzzing, mutation-based fuzzing, fuzzing in practice and the main differences between these projects. After that we summarize the inspirations and introduce our work.

\subsection{Generation-based Fuzzing}
Generation-based fuzzing generates a massive number of test cases according to  the specification of input format, e.g. a grammar. To fuzz the target applications that require inputs in complex format, the specifications used are crucial.
There are many types of specifications. Input model and context-free grammar are the two most common types.
Model-based fuzzers \cite{peach,skyfire,spike} follow a model of protocol. Hence, they are able to find more complex bugs by creating complex interactions with the target applications.
Peach \cite{peach} is one of the most popular model-based fuzzers with both generation and mutation abilities.
It develops two key models:
the data model determines the format of complex inputs and the state model describes the concrete method for cooperating with fuzzing targets.
By integrating fuzzing with models of data and state, Peach works effectively. 
Skyfire \cite{skyfire} first learns a context-sensitive grammar model, and then it generates massive inputs based on this model.

Some other popular fuzzers \cite{godefroid2008grammar,csmith,lava,ifuzzer,holler2012fuzzing} generate inputs based on context free grammar. 
P Godefroid \cite{godefroid2008grammar} enhances the whitebox fuzzing of complex structured-input applications by using symbolic execution, which directly generates grammar-based constraints whose satisfiability is examined using a custom grammar-based constraint solver.
Csmith \cite{csmith} is designed for fuzzing C-compilers. It generates plenty of random C programs in the C99 standard as the inputs. 
This tool can be used to generate C programs exploring a typical combination of C-language features while being free of undefined and unspecified behaviors. 
LAVA \cite{lava} generates effective test suites for the Java virtual machine by specifying production grammars.
IFuzzer \cite{ifuzzer} first constructs parse trees based on a language's context-free grammar, then it generates new code fragments according to these parse trees.
Radamsa \cite{helin2016radamsa} is a widely used generation-based fuzzer. It works by reading sample files of valid data and generating interestingly different outputs from them. Radamsa is an extreme "black-box" fuzzer, it needs no information about the program nor the format of the data. One can pair it with coverage analysis during testing to improve the quality of the sample set during a continuous fuzzing test.

\subsection{Mutation-based Fuzzing}
Mutation-based fuzzers \cite{zzuf,symfuzz,bff} mutate existing test cases to generate new test cases without any input grammar or input model specification.
Traditional mutation-based fuzzers such as zzuf \cite{zzuf} mutate the test cases by flipping random bits with a predefined ratio.
In contrast, the mutation ratio of SYMFUZZ \cite{symfuzz} is assigned dynamically. To detect bit dependencies of the input,
it leverages white-box symbolic analysis on an execution trace, then it dynamically computes an optimal mutation ratio according to these dependencies. Furthermore, BFF \cite{bff} integrates machine learning with evolutionary computation techniques to reassign the mutation ratio dynamically. 

Other popular AFL family tools \cite{afl,aflfast,aflgo,FairFuzz} apply various strategies to boost the fuzzing process. AFLFast \cite{aflfast} regards the process of target application as a Markov chain. A path-frequency based power schedule is responsible for computing the times of random mutation for each seed.
As with AFLFast, AFLGo \cite{aflgo} also proposes a simulated annealing-based power schedule, which helps fuzz the target code. %Especially, it describes a method to measure the distance between the seeds and the targets.
FairFuzz \cite{FairFuzz} mainly focuses on the mutation algorithm. It only mutates seeds that hit rare branches and it strives to ensure that the mutant seeds hit the rarest one.
(Wen Xu et.al.)\cite{xu2017designing} propose several new primitives ,
speeding up AFL by 6.1 to 28.9 times.
Unlike AFL family tools which track the hit count of each edge, libFuzzer \cite{libFuzzer} and honggfuzz \cite{Honggfuzz} utilize the SanitizerCoverage instrumentation method provided by the Clang compiler. To track block coverage, they track the hit count of each block as a guide to mutate the seeds during fuzzing.
SlowFuzz \cite{petsios2017slowfuzz} prioritizes seeds
that use more computer resources (e.g., CPU, memory and energy), increasing the probability of triggering algorithmic complexity
vulnerabilities. Furthermore, some fuzzers use concolic executors for hybrid fuzzing. Both Driller \cite{driller} and QSYM use mutation-based fuzzers to avoid path exploration of symbolic execution, while concolic execution is selectively used to drive execution across the paths that are guarded by narrow-ranged constraints.

\subsection{Cluster and Parallel Fuzzing in Industry}
Fuzzing has become a popular vulnerability discovery solution in industry \cite{liang2018fuzz} and has already found a large number of dangerous bugs and security vulnerabilities across a wide range of systems so far. For example, Google's OSS-Fuzz \cite{OSS-Fuzz} platform has found more than 1000 bugs in 5 months with thousands of virtual machines \cite{bug-report}.
% by using several popular fuzzers, including libFuzzer, honggfuzz \cite{Honggfuzz} and AFL. 
%In fact, the OSS-Fuzz platform is closely-related to the idea behind ensemble fuzzing to use multiple different fuzzers together. 
%However, it uses these fuzzers independently and does not combine them together effectively. In this way, the OSS-Fuzz platform can not improve the performance of these fuzzers themselves and has the additional problem of consuming significant computing resources. 
ClusterFuzz is the distributed fuzzing infrastructure behind OSS-Fuzz, and automatically executes libFuzzer powered fuzzer tests on scale \cite{ClusterFuzz,ClusterFuzz_two}. Initially built for fuzzing Chrome at scale, ClusterFuzz integrates multiple distributed libFuzzer processes, and performs effectively with corpus synchronization. 
ClusterFuzz mainly runs multiple identical instances of libFuzzer on distributed system for one target application. There is no diversity between these fuzzing instances.

%% 加一小段fuzzer的多线程说明，差异性不够
In industrial practice, many existing fuzzers also provide a parallel mode, and they work well with some synchronization mechisms.
%which allows fuzzing one target application with multiple parallel fuzzers across multiple cores. 
For example, each instance of AFL in parallel mode will periodically re-scan the top-level sync directory for any test cases found by other fuzzers\cite{AFLP1, AFLP2}. libFuzzer in parallel will also use multiple fuzzing engines to exchange the corpora\cite{libFuzzerP}.
These parallel mode can effectively improve the performance of fuzzer. 
%The more fuzzing jobs that are used, the better they perform. 
In fact, the parallel mode can be seen as a special example of ensemble fuzzing which uses multiple same base fuzzers. However, all these base fuzzers have a lack of diversity when using the same fuzzing strategy. %Consequently, only minor performance improvements are observed.

\begin{comment}
\subsubsection{Ensemble Learning}
Ensemble learning is a machine learning paradigm where multiple learners are trained to solve the same problem \cite{alpaydin2009introduction}. In contrast to ordinary machine learning approaches which try to learn one hypothesis from training data, ensemble methods try to construct a set of hypotheses and combine them for precision.
The diversity of base learners \cite{galar2012review} and the concrete ensemble methods \cite{zhou2012ensemble} are the two cores of ensemble learning. How to select base learners with great diversity and how to combine these base learners together are critical to the performance of ensemble learners. Many researches focus on these two parts to improve the generalization ability and the prediction accuracy of ensemble learners \cite{schapire1990strength, bauer1999empirical, ting1999issues, ting1999issues}. A number of diversity measures have been designed in ensemble learning \cite{krogh1995neural, kuncheva2003measures}. In addition, there are many effective ensemble methods, such as Boosting \cite{breiman1996bagging}, Bagging \cite{krogh1995neural, freund1997decision} and Stacking \cite{wolpert1992stacked}.
Ensemble learning has achieved a great success in most machine learning areas, contributing to better generalization ability.
\end{comment}

%% 总结一下
\subsubsection{Main Differences} Unlike the previous works, we are not proposing a new concrete generation-based or mutation-based fuzzing strategy. 
%or run different fuzzers on different machines, 
Nor do we run multiple identical fuzzers with multiple cores or machines. Instead, inspired by the seed synchronization of ClusterFuzz and AFL in parallel mode, we systematically study the possibility of the ensemble fuzzing of diverse fuzzers mentioned in the earlier works. Referred to the kernel descriptions of the evaluating fuzzing guidelines\cite{klees2018evaluating}, we empirically evaluate most state-of-the-art fuzzers, and identify some valuable results, especially for their performance variation across different real applications. To generate a stronger ensemble fuzzer, we choose multiple base fuzzers that are as diverse as possible based on three heuristics. We then implement an ensemble approach with global asynchronous and local synchronous based seed synchronization.

\section{Motivating Example}
\label{Motivating Example}
\begin{figure}[!htb]
 \centering
 \includegraphics[width=0.4\textwidth]{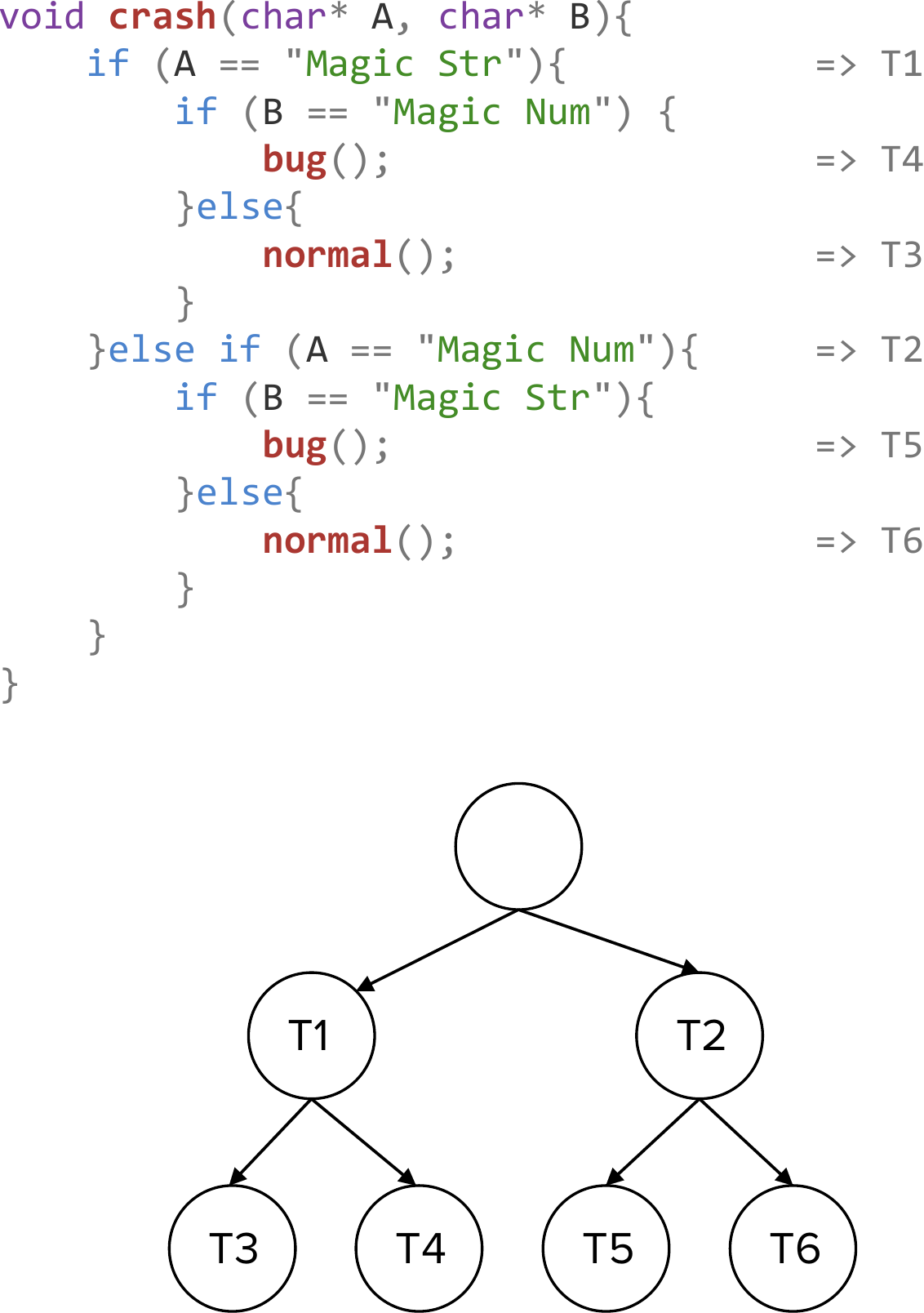}
 \caption{Motivating example of ensemble fuzzing with seed synchronization.}
 \label{fig:motivating example}
\end{figure}

	To investigate the effectiveness of ensemble fuzzing, we use a simple example in Figure \ref{fig:motivating example}
which takes two strings as input, and crashes when one of the two strings is ``Magic Str'' and the other string is ``Magic Num''.

%\begin{figure}[!htbp]
%\centering
%\begin{minipage}[!htbp]{0.4\textwidth}
%     \centering
%     \begin{minted}{c}
%void crash(char* A, char* B){
%    if (A == "Magic Str"){
%        if (B == "Magic Num") {
%            crash();
%        }else{
%            normal();
%        }
%    }else if (A == "Magic Num"){
%        if (B == "Magic Str"){
%            crash();
%        }else{
%            normal();
%        }
%    }
%}
%\end{minted}   
%     \small{(a)Code of Motivating example}
%\end{minipage}
%\begin{minipage}[!htbp]{0.40\textwidth}
%     \centering
%     \includegraphics[width=1.0\textwidth]{img/branch_example.pdf}     
%     \small{(b)Branches of Motivating example}
%\end{minipage}
%\caption{Motivating Example of Ensemble Fuzzing.}
%\label{fig:motivating example}
%\end{figure}

Many existing fuzzing strategies tend to be designed with certain preferences. %, and the coverage of these preferences varies greatly on the same application.
Suppose that we have two different fuzzers \(fuzzer_1\) and \(fuzzer_2\): \(fuzzer_1\) is good at solving the "Magic Str" problem, so it is better for reaching targets T1 and T3, but fails to reach targets T2 and T4. \(fuzzer_2\) is good at solving the "Magic Num" problem so it is better for reaching targets T2 and T6, but fails to reach targets T1 and T5.
If we use these two fuzzers separately, we can only cover one path and two branches.
At the same time, if we use them simultaneously and ensemble their final fuzzing results without seed synchronization, we can cover two paths and four branches.
However, if we ensemble these two fuzzers with some synchronization mechanisms throughout the fuzzing process, then, once \(fuzzer_1\) reaches T1, it synchronizes the seed that can cover T1 to \(fuzzer_2\). As a result, then, with the help of this synchronized seed, \(fuzzer_2\) can also reach T1, and because of its ability to  solve the "Magic Num" problem, \(fuzzer_2\) can further reach T4. Similarly, with the help of the seed input synchronized by \(fuzzer_2\), \(fuzzer_1\) can also further reach T2 and T5.
Accordingly, all four paths and all six branches can be covered through this ensemble approach.

\newcommand{\mysize}{\small}
\newcolumntype{C}{>{\arraybackslash}p{0.8cm}}
\NewEnviron{mytable_motivate}[2]{
  \begin{table}[!htbp]
  	\renewcommand\arraystretch{1.5}
    \caption{#1}
    \scalebox{0.9}[0.9]{%
      \label{tab:#2}
      \begin{tabular}{p{4.2cm}|CCCp{0.6cm}}
        \toprule
        {\mysize Tool}
        & {\mysize T1-T3}
        & {\mysize T1-T4}
        & {\mysize T2-T5}
        & {\mysize T2-T6}\\
        \midrule
        \BODY
        \bottomrule
      \end{tabular}
    }
  \end{table}%
}

\begin{mytable_motivate}{covered paths of each fuzzing option}{motivating_list} 
fuzzer1  &\checkmark \par   &   &   &   \\  
fuzzer2 &     &     &     &\checkmark \par    \\  
ensemble fuzzer1 and fuzzer2 without seed synchronization   & \checkmark \par  &    &    & \checkmark \par  \\  
ensemble fuzzer1 and fuzzer2 with seed synchronization       & \checkmark \par   & \checkmark \par   & \checkmark \par   & \checkmark \par   \\  
\end{mytable_motivate}

%过渡段，以上示例能够有效工作主要基于以下两点假设: 1,两个fuzzer是不同，(diverisity 比较大) 2. 在多个base fuzzer地fuzzing过程中存在高效地消息同步机制，能实时同步关键种子。 因此，ensmeble fuzzing 有两个重要地点： diversity地定义和挑选base fuzzer很重要 ensemble method (这段话非常关键！！！)
The ensemble approach in this motivating example works based on the following two hypotheses: (1) \(fuzzer_1\) and \(fuzzer_2\) expert in different domains; (2) the interesting seeds can be synchronized to all base fuzzers in a timely way.
To satisfy the above hypotheses as much as possible, successful ensemble fuzzers rely on two key points: (1) 
the first is to select base fuzzers with great diversity (as yet to be well-defined); 
(2) the second is a concrete synchronization mechanism to enhance effective cooperation among those base fuzzers.

\section{Ensemble Fuzzing}
\label{Ensemble Fuzzing Approach EnFuzz}
  %% 稍微较为详细的介绍一下集成学习，引出集成fuzz以及两个重要的点，多样性定义和集成的方式

\begin{figure}[!htb]
 \centering
 \includegraphics[width=0.47\textwidth]{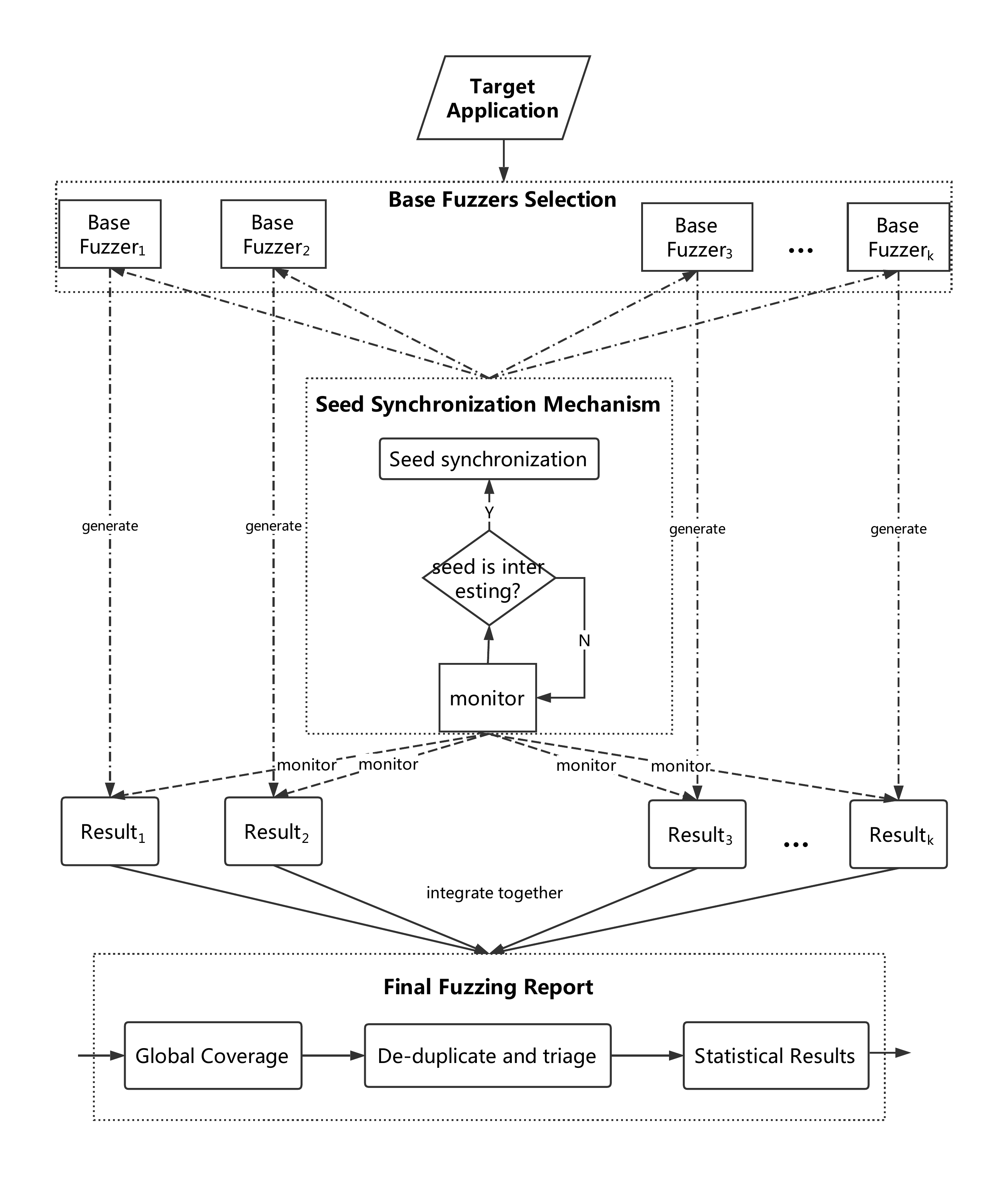}
 \vspace{-0.5 cm}
 \caption{The overview of ensemble fuzzing consists of base fuzzer selection and ensemble architecture design. The base fuzzer selection contains the diversity heuristic definition, and the architecture design includes the seed synchronization mechanism as well as final fuzzing report.}
 \vspace{-0.5 cm}
 \label{fig:framework_of_ensemble_fuzzing}
\end{figure}

%The main idea of 
For an ensemble fuzzing, we need to construct a set of base fuzzers and seamlessly combine them to test the same target application together. 
The overview of this approach is presented in Figure \ref{fig:framework_of_ensemble_fuzzing}. 
When a target application is prepared for fuzzing, we first choose several existing fuzzers as base fuzzers. 
%In fuzzing, the generalization refers to the ability of an fuzzing strategies to be effective across a range of target applications. 
%The generalization ability of a fuzzer can be described as the ability to perform well, no matter what the target application is.
The existing fuzzing strategies of any single fuzzer %have generalization limitations, because most of these strategies 
are usually designed with preferences. In real practice, these preferences vary greatly across different applications. They can be helpful in some applications, but may be less effective on other applications. Therefore, choosing base fuzzers with more diversity can lead to better ensemble performance. %We provide some initial quantification for their diversity. 
%The existing fuzzing strategies of any single fuzzer are usually designed with some preferences, and the performance of these preferences varies greatly on different applications. They can be helpful in some applications, but not certainly effective on other applications. 
After the base fuzzer selection, we integrate fuzzers with the globally asynchronous and locally synchronous based seed synchronization mechanism so as to monitor the fuzzing status of these base fuzzers and share interesting seeds among them. 
Finally, we collect crash and coverage information and feed this information into the fuzzing report.

 \vspace{-0.3cm}	
	\subsection{Base Fuzzer Selection}
	\label{Base Fuzzer Selection}
    %% 详细介绍base fuzzers 的多样性

The first step in ensemble fuzzing is to select a set of base fuzzers. These fuzzers can be generation-based fuzzers, e.g. Peach and Radamsa, or mutation-based fuzzers, e.g. libFuzzer and AFL. We can randomly choose some base fuzzers, but selecting base fuzzers with well-defined diversity improves the performance of an ensemble fuzzer. 

\begin{comment}
Let us take the coverage information for example. Suppose that we have a set of base fuzzers \(F = \{ fuzzer_1, fuzzer_2, \ldots , fuzzer_k \} \). The code coverage of target program \(P\) by \(fuzzer_j \) is \(C_j\). If we use all these \(k\) base fuzzers together to fuzz the target program \(P\), the total code coverage of \(P\) will be their union, that is \( C = C_1 \cup C_2 \cup C_3 \ldots \cup C_k \). The following two points are critical: 
%This is the core idea of the ensemble fuzzing.
%Furthermore, there are two key points we need to know:
\vspace{0.03in}
\begin{itemize}

	 \item [1.] If \( C_1 \cap C_2 \cap C_3 \ldots \cap C_k  \neq \emptyset \), then \( |C| \>> |C_j|\) where \(j\in 1, \ldots , k\), it means that the ensemble fuzzer always performs better than that of any base fuzzer.  

	 \item [2.] The smaller \( |C_i \cap C_j| \) is, the bigger \( |C_i \cup C_j|\) is, and the bigger \(|C|\) is, where \(0 \leq i, j \leq k, i \neq j \). 
	 
\end{itemize}
\vspace{0.03in} 
Therefore, the diversity of base fuzzers is the difference between their fuzzing strategies. The greater the difference of these base fuzzers, the more diversity they have, the better coverage ensemble fuzzers perform. 
%Here, for the mutation-based fuzzers, 
%We propose some simple solutions to these question based on the following three intuitions:
\end{comment}

We classify the diversity of base fuzzers according to three heuristics: seed mutation and selection strategy diversity, coverage information granularity diversity, inputs generation strategy diversity. The diversity heuristics are as follows: 

\begin{itemize}

\item  [1.] Seed mutation and selection strategy based heuristic: the diversity of base fuzzers can be determined by the variability of seed mutation strategies and seed selection strategies. %These two strategies are critical to mutation-based fuzzers.
%The main difference of most fuzzers are these fuzzing strategies. 
For example, AFLFast selects seeds that exercise low-frequency paths and mutates them more times, FairFuzz strives to ensure that the mutant seeds hit the rarest branches.

\item  [2.] Coverage information granularity based heuristic: many base fuzzers determine interesting inputs by tracking different coverage information. Hence, the coverage information is critical, and different kinds of coverage granularity tracked by fuzzers enhances diversity. For example, libFuzzer guides seed mutation by tracking block coverage while AFL tracks edge coverage.

\item  [3.] Input generation strategy based heuristic: fuzzers with different input generation strategies are suitable for different tasks. For example, generation-based fuzzers use the specification of input format to generate test cases, while the mutation-based fuzzers mutate initial seeds by tracking code coverage. So the generation-based fuzzers such as Radamsa perform better on complex format inputs and the mutation-based fuzzers such as AFL prefer complex logic processing.

\end{itemize}

Based on these three basic heuristics, we should be able to select a diverse set of base fuzzers with large diversity. It is our intuition that the diversity between the fuzzers following in two different heuristics is usually larger than the fuzzers that follows in the same heuristic. So, the diversity among the AFL family tools should be the least, while the diversity between Radamsa and AFL, between Libfuzzer and AFL, and between QSYM and AFL is should be greater. In this paper, we select base fuzzers manually based on the above heuristics. the base fuzzers will be dynamically selected according to the real-time coverage information.

	\subsection{Ensemble Architecture Design}
	\label{Ensemble Architecture Design}
    %% 详细介绍base fuzzers 的多样性

After choosing base fuzzers, we need to implement a suitable architecture to integrate them together. As presented in Figure \ref{fig:framework_of_ensemble_fuzzing}, 
%to make those fuzzers cooperate with each other,
inspired by the seed synchronization of AFL in parallel mode, 
one core mechanism is designed --- the globally asynchronous and locally synchronous (GALS) based seed synchronization mechanism. 
The main idea is to identify the interesting seeds (seeds that can cover new paths or new branches or can detect new unique crashes) from different base fuzzers asynchronously and share those interesting seeds synchronously among all fuzzing processes. %In the future, we will also support dynamic core allocation and different local synchronization periods adaption for different fuzzer instance with GALS design.  %This is consistent with the traditional GALS system design \cite{muttersbach2000practical}. 
%Here, we propose two core ensemble mechanism: 

%\vspace{0 cm}
\begin{figure}[!htb]
 \centering
 \includegraphics[width=0.5\textwidth]{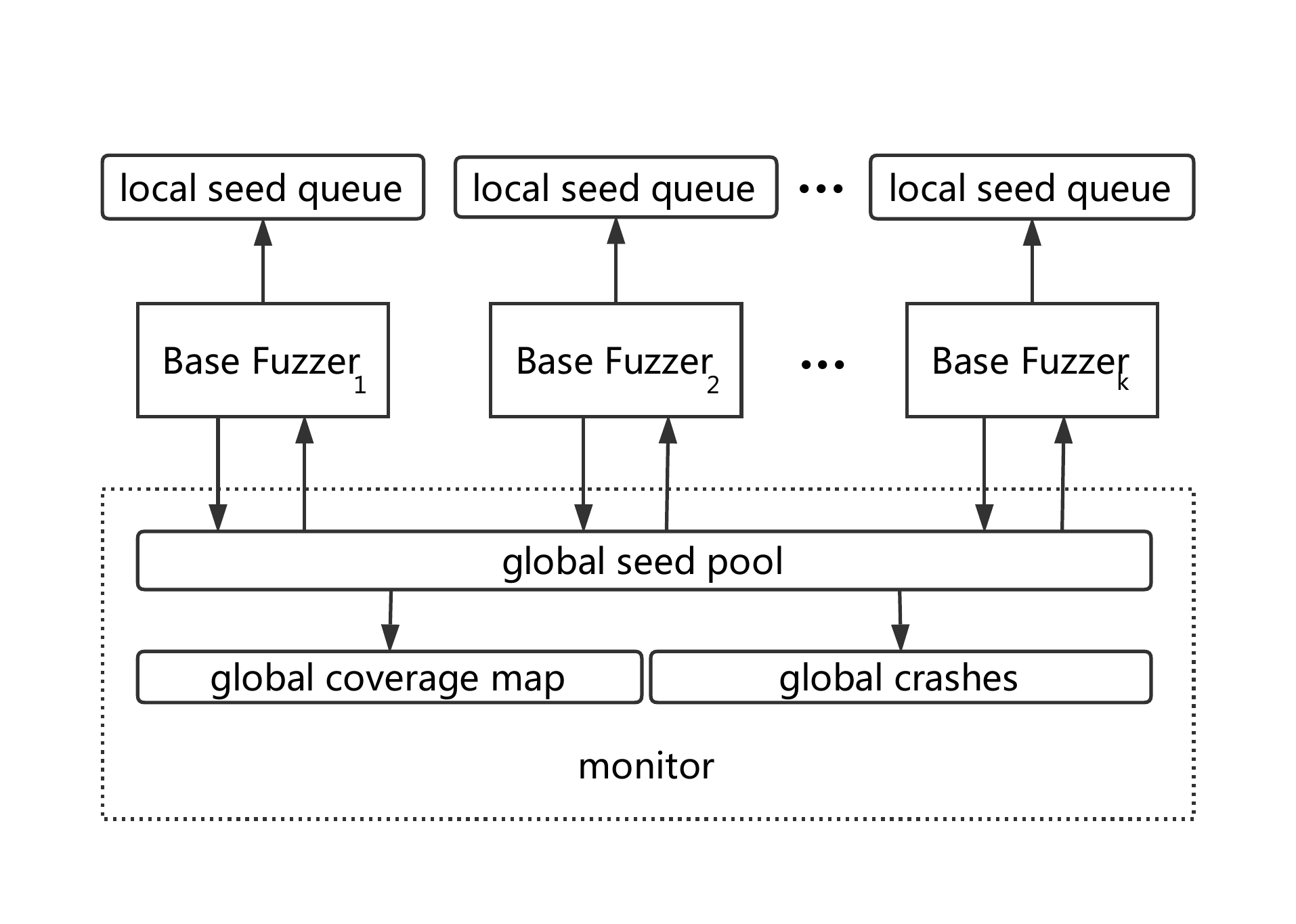}
 \vspace{-1.5 cm}
 \caption{The data  structure of global asynchronous and local synchronous based seed synchronization mechanism.}
 \label{fig:ensemble_architecture}
\end{figure}

\IncMargin{1em}
\begin{algorithm}
\SetAlgoLined
\SetKwData{Left}{left}
\SetKwFunction{Mutate}{Mutate}
\SetKwFunction{Run}{Run}
\SetKwFunction{push}{push}

\SetKwFunction{causeCrash}{causeCrash}
\SetKwFunction{haveNewCoverage}{haveNewCoverage}
\SetKw{Continue}{continue}
\SetKwInOut{Input}{Input}\SetKwInOut{Output}{Output}

\Input{Local seed pool of base fuzzer $queue$}

%\BlankLine

\Repeat{timeout or abort-signal}{
    \ForEach{\textrm{seed} $s$ \textrm{of the} queue}{\
    	$s'=\Mutate{s}$\;
    	$Cover=\Run{s'}$\;\
    	{\color{gray} // if seeds $s'$ causes new crash or have new \newline
    				// coverage, then store it in own seed pool and \newline
    				// push it to the global seed pool asynchronously\;}
        \uIf{$Cover$.\causeCrash{}}{
           $crashes.\push{s'}$\;
           $queue.\push(s')$\;
           $GlobalSeedPool.\push(s')$\;
        }
        \ElseIf{$Cover$.\haveNewCoverage{}} {
           $queue.\push(s')$\;
           $GlobalSeedPool.\push(s')$\;
        }
    }
}
\Output{Global crashing seeds $crashes$}
\caption{Action of local base fuzzer}\label{sync_baseFuzzer}
\end{algorithm}\DecMargin{1em}

This seed synchronization mechanism employs a global-local style data structure as shown in Figure \ref{fig:ensemble_architecture}. The local seed queue is maintained by each base fuzzer, while the global pool is maintained by the monitor for sharing interesting seeds among all base fuzzers. In ensemble fuzzing, the union of these base fuzzers' results is needed to identify interesting seeds during the whole fuzzing process. Accordingly, the global coverage map is designed, and any new paths or new branches covered by the interesting seeds will be added into this global map. This global map can not only help decide which seeds to be synchronized, but also help de-duplicate and triage the results. Furthermore, to output the final fuzzing report after completing all fuzzing jobs, any interesting seeds which contribute to triggering unique crashes will be stored in the global crashes list.

First, let us take a look at the seed synchronization solution of the base fuzzer, which mainly describes how base fuzzers contribute the interesting seeds asynchronously to the global pool. As presented in lines 2-4 of algorithm \ref{sync_baseFuzzer}, 
for each single base fuzzer, it works with a local input seed queue and runs a traditional continuous fuzzing loop. It has three main steps: (1) Select input seeds from the queue, (2) mutate the selected input seeds to generate new candidate seeds, (3) run the target program with the candidate seeds, track the coverage and report vulnerabilities. Once the candidate seeds have new coverage or cause unique crashes, they will be regarded as interesting seeds and be pushed asynchronously into the global seed pool, as presented in lines 6-12. %of Algorithm \ref{sync_baseFuzzer}. 

\vspace{-0.2cm}
\IncMargin{1em}
\begin{algorithm}
\SetAlgoLined
\SetKwData{Left}{left}
\SetKwFunction{setup}{setup}
\SetKwFunction{initial}{initial}
\SetKwFunction{push}{push}
\SetKwFunction{isSync}{isSync}
\SetKwFunction{setSync}{setSync}
\SetKwFunction{sleep}{sleep}
\SetKwFunction{run}{run}
\SetKwFunction{isEmpty}{isEmpty}
\SetKwFunction{causeCrash}{causeCrash}
\SetKw{Continue}{continue}
\SetKwInOut{Input}{Input}\SetKwInOut{Output}{Output}

\Input{Base fuzzers list $BaseFuzzers[]$ \newline
		Initial seeds $S$ \newline
		Synchronization period $period$}

%\BlankLine
\
{\color{gray} // set up each base fuzzers \;}
\ForEach{\textrm{base fuzzer} $f$ \textrm{of the} $BaseFuzzers[]$}{\
		$fuzzer.\setup{}$\;
}\
{\color{gray} // set up thread monitor for monitoring \;}
$monintor.\setup()$\;
$GlobalCover.\initial{}$\;
$GlobalSeedPool.\initial{}$\;
$GlobalSeedPool.\push{S}$\;
\Repeat{timeout or abort-signal}{
    \ForEach{\textrm{seed} $s$ \textrm{of the} $GlobalSeedPool$}{\
    	{\color{gray} // Skip synchronized seeds \;}
        \uIf{ $s.\isSync{}$ == $False$ }{
        	\ForEach{\textrm{base fuzzer} $f$ \textrm{of the} $BaseFuzzers[]$}{\
			 	$Cover=f.\run{s}$ \;\ 	
			 	{\color{gray} // update the global coverage \;}\
				$newCover=(Cover \cup GlobalCover)-GlobalCover$ \;\
				$GlobalCover=Cover \cup GlobalCover$\;\
				{\color{gray} // synchronize the seed $s$ to base fuzzer $f$ \;}\
				\uIf{$Cover$.\causeCrash{} and $!newCover.\isEmpty{}$}{
					$crashes.\push{s}$\;
					$f.queue.\push{s}$\;
				}\uElseIf{$!newCover.\isEmpty{}$ }{
					$f.queue.\push{s}$\;
				}
				\Else {
            		\Continue\;
        		}
			}
        }
        \Else {
            \Continue\;
        }
        $s.\setSync{True}$\;
    }\
    {\color{gray} // waiting until next seed synchronization \;}
    $\sleep{period}$\;
}
\Output{Crashing seeds $crashes$}
\caption{Action of global monitor $sync$}\label{sync_monitor}
\end{algorithm}\DecMargin{1em}

Second, let us see the seed synchronization solution of the monitor process, which mainly describes how the monitor process synchronously dispatches the interesting seeds in the global pool to the local queue of each base fuzzer.
When all base fuzzers are established, a thread named \texttt{monitor} will be created for monitoring the execution status of these fuzzing jobs, as in lines 2-6 of algorithm \ref{sync_monitor}. It initializes the global coverage information to record the global fuzzing status of target applications by all the base fuzzer instances and then creates the global seed pool with the initial seeds, as in lines 7-9 of algorithm \ref{sync_monitor}. It then runs a continuous periodically synchronizing loop --- each base fuzzer will be synchronously dispatched with the interesting seeds from the global seed pool. Each base fuzzer will incorporate the seeds into its own local seed queue, once the seeds are deemed to be interesting seeds (seeds contribute to the coverage or crash and has not been generated by the local fuzzer), as in line 15-24 .
To lower the overhead of seed synchronization, a thread \texttt{monitor} is designed to work periodically. Due to this globally asynchronous and locally synchronous based seed synchronization mechanism, base fuzzers cooperate effectively with each other as in the motivating example in Figure 1.

%\subsubsection{Seed Synchronization Mechanism}
%As mentioned in section \ref{Related Work}, in machine learning, researchers only ensemble base learners with their final results; but in fuzzing, we focus on the union of coverage throughout the fuzzing process. Consequently, to enhance cooperation among these base fuzzers during their fuzzing processes, the seed synchronization mechanism is designed. %, as presented in Figure \ref{fig:ensemble_architecture}.

%Unlike ensemble learning which only wants averaging or voting of final results, ensemble fuzzing wants the union of base fuzzers' results. Accordingly, the global coverage map is designed to solve this problem, as presented in line 17-18 of Algorithm \ref{sync_monitor}. Any interesting seeds which contribute to cover new branches or paths, or trigger new crashes will be union into the global coverage map. This global map not only help decide which seeds to be synchronized, but also help de-duplicate and triage the results, output the final fuzzing report after the completion of all fuzzing jobs.
%
%
%
%The seed synchronization mechanism works with the global coverage map to make these base fuzzers cooperate with each other deeply and effectively. 
%There are many other effective and advanced ensemble methods in ensemble learning, such as Boosting, which could also be applied and customized based on this architecture.

\section{Evaluation}
\label{Evaluation}
To present the effectiveness of ensemble fuzzing, we first implement several prototypes of ensemble fuzzer based on the state-of-the-art fuzzers. Then, we refer to some kernel descriptions of evaluating fuzzing guideline \cite{klees2018evaluating}. We conduct thorough evaluations repeatedly   on LAVA-M and Google's fuzzer-test-suite, several well-fuzzed open-source projects from GitHub, and several commercial products from companies. Finally, according to the results, we answer the following three questions:
(1) Can ensemble fuzzer perform better?
(2) How do different base fuzzers affect Enfuzz?
(3) How does Enfuzz perform on real-world applications
%\textbf{(1) Can ensemble fuzzer perform better?
%(2) How do different base fuzzers affect Enfuzz?
%(3) How does Enfuzz perform on real-world applications?}

%\begin{itemize}

%	 \item [1.] Does Generalization Limitation exists in these base %fuzzers?

%	 \item [2.] How does diversity affect the performance of base fuzzers?

%	 \item [3.] Can ensemble fuzzing perform better with more diversity?
	 
%\end{itemize}

	\vspace{-0.2cm}
  \subsection{Ensemble Fuzzer Implementation}
  	\label{implementation}
    We implement ensemble fuzzing based on six state-of-the-art fuzzers, including three edge-coverage guided mutation-based fuzzers -- AFL, AFLFast and FairFuzz, one block-coverage guided mutation-based fuzzer -- libFuzzer, one generation-based fuzzer -- Radamsa and one most recently hybrid fuzzer -- QSYM. These are chosen as the base fuzzers for the following reasons (Note that \toolThree ~is not limited to these six and other fuzzers can also be easily integrated, such as honggfuzz, ClusterFuzzer etc.):

\begin{itemize}
\item Easy integration.
All the fuzzers are open-source and have their core algorithms implemented precisely. It is easy to integrate those existing fuzzers into our ensemble architecture. We do not have to implement them on our own, which eliminates any implementation errors or deviations that might be introduced by us.

\item Fair comparison. All the fuzzers perform very well and are the latest and widely used fuzzers, as is seen by their comparisons with each other in prior literature, for example, QSYM outperforms similar fuzzers such as Angora\cite{chen2018angora} and VUzzer. We can evaluate their performance on real-world applications without modification.

\item Diversity demonstration. All these fuzzers have different fuzzing strategies and reflect the diversity among  correspondence with the three base diversity heuristics mentioned in section \ref{Base Fuzzer Selection}: coverage information granularity diversity, input generation strategy diversity,  seed mutation and selection strategy diversity. The concrete diversity among these base fuzzers is listed in Table \ref{tab:diversity}.

\end{itemize}

%Several engineering challenges need to be solved firstly: 
To demonstrate the performance of ensemble fuzzing and the influence of diversity among base fuzzers, 
five prototypes are developed. (1) \toolOne, an ensemble fuzzer only based on AFL, AFLFast and FairFuzz. 
% thus the diversity of \toolOne ~is little. 
(2) \toolFive, an ensemble fuzzer based on AFL, AFLFast, FairFuzz and QSYM, a practical concolic execution engine is included. 
(3) \toolTwo, an ensemble fuzzer based on AFL, AFLFast, FairFuzz and libFuzzer, a block-coverage guided fuzzer is included. %, and the diversity of \toolTwo ~is larger. 
(4) \toolThree, an ensemble fuzzer based on AFL, AFLFast, libFuzzer and Radamsa, a generation-based fuzzer is further added .%, and the diversity of \toolThree ~is the largest among them.
(5) \toolFour, ~with the ensemble of same base fuzzers (AFL, AFLFast and FairFuzz), but without the seed synchronization, to demonstrate the effectiveness of the global asynchronous and local synchronous based seed synchronization mechanism.  
During implementation of the proposed ensemble mechanism, we address the following challenges:

\newcolumntype{C}{>{\arraybackslash}p{6.3cm}}
\NewEnviron{mytable_diversity}[2]{
  \begin{table}[tbp]
    \caption{#1}
    \scalebox{1.0}[1.0]{%
      \label{tab:#2}
      \begin{tabular}{l|Cp{1cm}}
        \toprule
        {\mysize Tool}
        & {\mysize diversity compared with AFL}\\
        \midrule
        \BODY
        \bottomrule
      \end{tabular}
    }
  \end{table}%
}

\begin{mytable_diversity}{Diversity among these base fuzzers}{diversity} 
AFLFast  		& Seed mutation and selection strategy based rule: the times of random mutation for each seed is computed by a Markov chain model. The seed selection strategy is different. \\	\hline 
FairFuzz 		& Seed mutation and selection strategy based rule: only mutates seeds which hit rare branches and strives to ensure the mutant seeds hit the rarest one. The seed mutation strategy is different. \\  \hline
libFuzzer  	 	& Coverage information granularity based rule: libFuzzer mutates seeds by utilizing the SanitizerCoverage instrumentation, which supports tracking block coverage; while AFL uses static instrumentation with a bitmap to track edge coverage. The coverage information granularity is different.\\	\hline
Radamsa       	& Input generation strategy based rule: Radamsa is a widely used generation-based fuzzer which generates different inputs sample files of valid data. The input generation strategy is different. \\		\hline
QSYM       	& QSYM is a practical fast concolic execution engine tailored for hybrid fuzzing. It makes hybrid fuzzing scalable enough to test complex, real-world
applications.   \\ % and outperforms similar fuzzers such as \textbf{Angora}
\end{mytable_diversity}

\textit{1) Standard Interface Encapsulating}
The interfaces of these fuzzers are different. For example, AFL family tools use the function \texttt{main}, but libFuzzer use a function \texttt{LLVMFuzzerTestOneInput}. Therefore, it is hard to ensemble them together.
We design a standard interface to encapsulate the complexity of different fuzzing tools.
This standard interface takes seeds from the file system, and writes the results back to the file system.
All base fuzzers receive inputs and produce results through this standard interface, through which different base fuzzers can be ensembled easily.

\textit{2) libFuzzer Continuously Fuzzing}
The fuzzing engine of libFuzzer  will be shut down when it finds a crash, while other tools continue fuzzing until manually closed. It is unfair to compare libFuzzer  with other tools when the fuzzing time is different.
The persistent mode of AFL is a good solution to this problem. Once AFL sets up, the fuzzer parent will \texttt{fork} and \texttt{execve} a new process to fuzz the target. When the target process crashes, the parent will collect the crash and resume the target, then the process simply loops back to the start.
Inspired by the AFL persistent mode, we set up a thread named \texttt{Parent} to monitor the state of libFuzzer. Once it shuts down, \texttt{Parent} will resume the libFuzzer. %and restart the fuzzing process.

\begin{comment}
\subsubsection{Radamsa Effectively Working}
Radamsa is a widely used generation-based fuzzer. There are two key obstacles to using Radamsa effectively. 
Firstly, Radamsa requires some sample files of input data, but AFL family tools and libFuzzer do not need. 
Secondly, AFL and libFuzzer mutate seeds by coverage information, but Radamsa does not, to compare them in terms of coverage is unfair. %It is unfair to compare generation-based fuzzers with mutation-based fuzzers directly in terms of path or branch coverage.
However, according to the original experiments of Radamsa and Skyfire, they pair generation-based fuzzers with mutation-based fuzzers to solve the two problems. We follow their setup to pare Radamsa with AFL: when AFL is running, Radamsa periodically read seeds generated by AFL as inputs and generates lots of outputs which will be fed back to AFL. Radamsa and AFL are running alternately on one CPU core.
%In this way, Radamsa will work effectively with AFL, and achieve high performance.
\end{comment}

\textit{3) Bugs De-duplicating and Triaging}
%To analyse crashes effectively, we need de-duplicate and triage these crashes first.
%Accordingly, 
We develop a tool for crash analysis.
We compile all the target applications with AddressSanitizer, and test them with the crash samples. 
When the target applications crash, the coredump file, which consists of the recorded state of the working memory will be saved. %Some key pieces of program state are usually dumped at the same time, including the processor registers, which may include the program counter and stack pointer, memory management information, and operating system flags and information.
Our tool first loads coredump files, then gathers the frames of each crash; finally, it identifies two crashes as identical if and only if the top frame is identical to the other frame. The method above is prone to underestimating bugs. For example, two occurrences of heap overflow may crash at the cleanup function at exit. However, the target program is instrumented with AddressSanitizer. As the program terminates immediately when memory safety problems occur, the top frame is always relevant to the real bug.
%Our tool automatically collects these coredump files, analyses and records each crash's stack backtraces based on LLDB, then de-duplicates and triages crashes according to these key information. %Finally, it reports the root causes of these crashes.
In practice, the original duplicate unique crashes have been drastically de-duplicated to a humanly check-able number of unique bugs, usually without duplication. Even though there are some extreme cases that different top frames for one bug, the result can be further refined by manual crash analysis.

\textit{4) Seeds effectively Synchronizing}
The implementation of the seed synchronization mechanism: 
all base fuzzers have implemented the communication logic following the standard interface. Each base fuzzer will put interesting seeds into its own local seed pool, and the monitor thread \texttt{sync} will periodically make each single base fuzzer pull synchronized seeds from the global seed pool through a communication channel. This communication channel is implemented based on file system. 
%The time period of the seed synchronization mechanism is set to 120 seconds. 
A shorter period consumes too many resources, which leads to a decrease in fuzzing performance. 
A longer period will make seed synchronizing untimely, which also affects the performance.
After multiple attempts with different values, it is found that the synchronization interval affects the performance at the beginning of fuzzing, while little impact was observed in the long term. The interval of 120s is identified with the fastest convergence.

%The implementation of result integration mechanism:
%the core idea of result integrating is de-duplicating the paths covered, branches exercised and crashes detected by all the base fuzzers.
%The crash de-duplicating has been introduced above; similar to it, to de-duplicate the paths and branches, we develop a tool to collect all seeds generated by each base fuzzer into a global seeds pool automatically. Then our tool runs the target application with these seeds one by one, at the meantime the tool will analyse and record paths and branches coverage of each seed. After processing all seeds, the global coverage information can be reported.

  \vspace{-0.1cm}
  \subsection{Data and Environment Setup}
    Firstly, we evaluate ensemble fuzzing on LAVA-M \cite{dolan2016lava}, which consistis of four buggy programs, file, base64, md5sum and who.  LAVA-M is a test suite that injects hard-to-find bugs in Linux utilities to evaluate bug-finding techniques. Thus the test is adequate for demonstrating the effectiveness of ensemble fuzzing.
Furthermore, to reveal the practical performance of ensemble fuzzing, we also evaluate our work based on fuzzer-test-suite \cite{fuzzer_test_suite}, a widely used benchmark from Google. The test suite consists of popular open-source real-world applications.
%including well-known tools (e.g. libarchive, sqlite), 
%image processing libraries (e.g. guetzli, libjpeg, libpng),  
%communication protocol toolkits  (e.g. openssl, boringssl), 
%color management engines (e.g. lcms),
%text engines (e.g. freetype2, harfbuzz),
%regular expression engines (e.g. proj4, re2),
%and document processers (e.g. json, libxml2), etc.
This benchmark is chosen to avoid the potential bias of the cases presented in literature, and for its great diversity, which helps demonstrate the performance variation of existing base fuzzers.

We refer to the kernel criteria and settings of evaluation from the fuzzing guidelines \cite{klees2018evaluating}, and integrate the three widely used metrics from previous literature studies to compare the results on these real-world applications more fairly, including the number of paths, branches and unique bugs. 
%As described in \cite{klees2018evaluating}, the number of unique crashes  can overestimate the number of bugs significantly, so we choose to unique bugs.
To get unique bugs, we use crash's stack backtraces to deduplicate unique crashes, as mentioned in the previous subsection. % as described in \cite{klees2018evaluating}. 
%Note that AFL-based fuzzers distinguish the crash by execution paths, therefore, some crashes could be caused by an identical root cause. But in general, the more crashes we find, the higher probability that more vulnerabilities can be identified. The number of unique crashes indicates the probability of finding vulnerabilities of target applications. Even we can get the better ability indication metrics such as unique bugs in the future as described in \cite{klees2018evaluating}, the unique crash might be kept as an important metric. 
The initial seeds for all experiments are the same. We use the test cases originally included in their applications or empty seed if such initial seeds do not exist. %The first two metrics can evaluate the coverage of the target applications, as coverage is one of the most important factors contributing to the success of fuzzers.
%In addition to coverage information, another important factor to fuzzers is the number of unique crashes detected. 

The experiment on fuzzer-test-suite is conducted ten times in a 64-bit machine with 36 cores (Intel(R) Xeon(R) CPU E5-2630 v3 @ 2.40GHz), 128GB of main memory, and Ubuntu 16.04 as the host OS with SMT enabled. Each binary is hardened by AddressSanitizer \cite{SanitizerCoverage} to detect latent bugs.
First, we run each base fuzzer for 24 hours with one CPU core in single mode.
Next, since \toolTwo, ~\toolThree ~and \toolFive ~need at least four CPU cores to ensemble these four base fuzzers, we also run each base fuzzer in parallel mode for 24 hours with four CPU cores. 
In particular, \toolOne ~and \toolFour ~only ensembles three types of base fuzzers (AFL, AFLFast and FairFuzz). To use the same resources, we set up two AFL instances, one AFLFast instance and one FairFuzz instance.
This experimental setup ensures that the computing resources usage of each ensemble fuzzer is the same as any base fuzzers running in parallel mode.
Due to the large amount of data and the page limitation, we include the variation of those statistical test results in our GitHub. In fact. most metrics converged to similar values during multithreaded fuzzing. The variation of those statistical test results is small (between -5\% ~ 5\%), so we just use the averages in this paper.

  \vspace{-0.1cm}
  \subsection{Preliminary Evaluation on LAVA-M }
    We first evaluate ensemble fuzzing on LAVA-M, which has been used for testing other systems such as Angora, T-Fuzz and QSYM, and QSYM shows the best performance. 
We run \toolFive ~(which ensembles AFL, AFLFast, FairFuzz and QSYM) on the LAVA-M dataset. %for five hours, which is the test duration set by the original LAVA work \cite{dolan2016lava}. 
To demonstrate its effectiveness, we also run each base fuzzer using the same resources --- four instances of AFL in parallel mode, four instances of AFLFast in parallel mode, four instances of FairFuzz in parallel mode, QSYM with four CPU cores used in parallel mode (two instances of concolic execution engine and two instances of AFL). To identify unique bugs, we used built-in bug identifiers provided by the LAVA project.
The results are presented in Table \ref{tab:lava_path},  \ref{tab:lava_branch} and  \ref{tab:lava_bug}, which show the number of paths executed, branches covered and unique bugs detected by AFL, AFLFast, FairFuzz, QSYM, \toolFive. 

From Tables \ref{tab:lava_path},  \ref{tab:lava_branch} and  \ref{tab:lava_bug}, we found that AFL, AFLFast and FairFuzz perform worse due to the complexity of their branches.
The practical concolic execution engine helps QSYM solve complex branches and find significantly more bugs. The base code of the four applications in LAVA-M are small (2K-4K LOCs) and concolic execution could work well on them. However, real projects have code bases that easily reach 10k LOCs. Concolic execution might perform worse or even get hanged, as presented in the latter subsections. 
Furthermore, when we ensemble AFL, AFLFast, FairFuzz and QSYM together with the GALS based seed synchronization mechanism -- \toolFive ~ always performs the best in both coverage and bug detection.
In total, compared with AFL, AFLFast, FairFuzz and QYSM, \toolFive ~ executes 44\%, 45\%, 43\% and 7.7\% more paths, covers 195\%, 215\%, 194\% and 5.8\% more branches, and detectes 8314\%, 19533\%, 12989\% and 0.68\% more unique bugs respectively. 
From these preliminary statistics, we can determine that the performance of fuzzers can be improved by our ensemble approach.

\newcolumntype{C}{R{1.1 cm}}
\NewEnviron{mytable_lava1}[2]{
  \begin{table}[!htbp]
    \caption{#1}
    \scalebox{0.9}[0.9]{%
      \label{tab:#2}
      \begin{tabular}{l|CCCCC}
        \toprule
        {\mysize Project}
        & {\mysize AFL}
        & {\mysize AFLFast}
        & {\mysize FairFuzz}
        & {\mysize QSYM}
        & {\mysize \toolFive}\\
        \midrule
        \BODY
        \bottomrule
      \end{tabular}
    }
  \end{table}%
}

%\vspace{-0.4cm}
\begin{mytable_lava1}{Number of paths covered by AFL, AFLFast, FairFuzz, QSYM and \toolFive ~on LAVA-M.}{lava_path}
base64 & 1078 & 1065 & 1080 & 1643 & \textbf{1794} \\
md5sum & 589  & 589  & 601  & 1062 & \textbf{1198} \\
who    & 4599 & 4585 & 4593 & 5621 & \textbf{5986} \\
uniq   & 476  & 453  & 471  & 693  & \textbf{731 } \\
\midrule 
total  & 6742 & 6692 & 6745 & 9019 & \textbf{9709} \\
\end{mytable_lava1}
%\vspace{-0.4cm}

%\vspace{-0.4cm}
\begin{mytable_lava1}{Number of branches covered by AFL, AFLFast, FairFuzz, QSYM and \toolFive ~on LAVA-M.}{lava_branch}
base64 & 388  & 358  & 389  & 960  & \textbf{993 } \\
md5sum & 230  & 208  & 241  & 2591 & \textbf{2786} \\
who    & 813  & 791  & 811  & 1776 & \textbf{1869} \\
uniq   & 1085 & 992  & 1079 & 1673 & \textbf{1761} \\
\midrule  
total  & 2516 & 2349 & 2520 & 7000 & \textbf{7409} \\
\end{mytable_lava1}
%\vspace{-0.4cm}

%\vspace{-0.4cm}
\begin{mytable_lava1}{Number of bugs found by AFL, AFLFast, FairFuzz, QSYM and \toolFive ~on LAVA-M.}{lava_bug}
base64 & 1  & 1 & 0 & 41            & \textbf{42  } \\
md5sum & 0  & 0 & 1 & \textbf{57}   & \textbf{57  } \\
who    & 2  & 0 & 1 & 1047          & \textbf{1053} \\
uniq   & 11 & 5 & 7 & 25            & \textbf{26  } \\
\midrule
total  & 14 & 6 & 9 & 1170         & \textbf{1178} \\
\end{mytable_lava1}

  \subsection{Evaluation on Google's fuzzer-test-suite}
     While LAVA-M is widely used, Google's fuzzer-test-suite is more practical with many more code lines and containing real-world bugs. To reveal the effectiveness of ensemble fuzzing, we run \toolThree ~(which only ensembles AFL, AFLFast, LibFuzzer and Radamsa) on all of the 24 real-world applications of Google's fuzzer-test-suite for 24 hours 10 times. %The reason we do not include QSYM will be discussed in \ref{Evaluation}. 
As a comparison, we also run each base fuzzer in parallel mode with four CPU cores used. To identify unique bugs, we used stack backtraces to deduplicate crashes.
The results are presented in Tables \ref{tab:google_path}, \ref{tab:google_branch} and \ref{tab:google_bug}, which shows the average number of paths executed, branches covered and unique bugs detected by AFL, AFLFast, FairFuzz, LibFuzzer, Radamsa, QSYM and \toolThree ~respectively.

\newcolumntype{B}{R{1.0 cm}}
\newcolumntype{C}{R{0.8 cm}}
\NewEnviron{mytable_google}[2]{
  \begin{table}[!htbp]
    \caption{#1}
    \scalebox{0.74}[0.74]{%
      \label{tab:#2}
      \begin{tabular}{l|CBCBCCB}
        \toprule
        {\mysize Project}
        & {\mysize AFL}
        & {\mysize AFLFast}
        & {\mysize FairFuzz}
        & {\mysize LibFuzzer}
        & {\mysize Radamsa}
        & {\mysize QSYM}
        & {\mysize \toolThree}\\
        \midrule
        \BODY
        \bottomrule
      \end{tabular}
    }
  \end{table}%
}

\vspace{-0.5cm}
\begin{mytable_google}{Average number of paths covered by each tool on Google's fuzzer-test-suite for ten times.}{google_path}
boringssl     & 3286   & 2816   & 3393   & 5525   & 3430   & 2973   & \textbf{7136 }  \\
c-ares        & 146    & 116    & 146    & 191    & 146    & 132    & \textbf{253  }  \\
guetzli       & 3248   & 2550   & 1818   & 3844   & 3342   & 2981   & \textbf{4508 }  \\
lcms          & 1682   & 1393   & 1491   & 1121   & 1416   & 1552   & \textbf{2433 }  \\
libarchive    & 12842  & 10111  & 12594  & 22597  & 12953  & 11984  & \textbf{31778}  \\
libssh        & 110    & 102    & 110    & 362    & 110    & 149    & \textbf{377  }  \\
libxml2       & 14888  & 13804  & 14498  & 28797  & 17360  & 13172  & \textbf{35983}  \\
openssl-1.0.1 & 3992   & 3501   & 3914   & 2298   & 3719   & 3880   & \textbf{4552 }  \\
openssl-1.0.2 & 4090   & 3425   & 3956   & 2304   & 3328   & 3243   & \textbf{4991 }  \\
openssl-1.1.0 & 4051   & 3992   & 4052   & 2638   & 3593   & 4012   & \textbf{4801 }  \\
pcre2         & 79581  & 66894  & 71671  & 59616  & 78347  & 60348  & \textbf{85386}  \\
proj4         & 342    & 302    & 322    & 509    & 341    & 323    & \textbf{709  }  \\
re2           & 12093  & 10863  & 12085  & 15682  & 12182  & 10492  & \textbf{17155}  \\
woff2         & 23     & 16     & 20     & 447    & 22     & 24     & \textbf{1324 }  \\
freetype2     & 19086  & 18401  & 20655  & 25621  & 18609  & 17707  & \textbf{27812}  \\
harfbuzz      & 12398  & 11141  & 14381  & 16771  & 11021  & 12557  & \textbf{16894}  \\
json          & 1096   & 963    & 721    & 1081   & 1206   & 1184   & \textbf{1298 }  \\
libjpeg       & 1805   & 1579   & 2482   & 1486   & 1632   & 1636   & \textbf{2638 }  \\
libpng        & 582    & 568    & 587    & 586    & 547    & 606    & \textbf{781  }  \\
llvm          & 8302   & 8640   & 9509   & 10169  & 8019   & 7040   & \textbf{10935}  \\
openthread    & 268    & 213    & 230    & 1429   & 266    & 365    & \textbf{1506 }  \\
sqlite        & 298    & 322    & 294    & 580    & 413    & 300    & \textbf{636  }  \\
vorbis        & 1484   & 1548   & 1593   & 1039   & 1381   & 1496   & \textbf{1699 }  \\
wpantund      & 4914   & 5112   & 5691   & 4881   & 4891   & 4941   & \textbf{5823 }  \\
\midrule  \textbf{
Tota}l         & 190607 & 168372 & 186213 & 209574 & 188274 & 163097 & \textbf{271408} \\
\midrule
Improvement &--     & \small{11\% $\downarrow$} & 2\% $\downarrow$ & 9\% $\uparrow$ & 1\% $\downarrow$ & 14\%$\downarrow$ & \textbf{42\% $\uparrow$} \\
\end{mytable_google}
\vspace{-0.2cm}

The first six columns of Table \ref{tab:google_path} reveal the issue of the performance variation in those base fuzzers, as they perform variously on different applications. Comparing
AFL family tools, AFL performs better than the other two optimized fuzzers on 14 applications. %FairFuzz performs the best on the other 2 applications, AFLFast fails to perform the best on any application. 
Compared with AFL, libFuzzer performs better on 15 applications, but worse on 9 applications. 
Radamsa performs better on 8 applications, but also worse on 16 applications.
QSYM performs better on 9 applications, but also worse on 15 applications.
Table \ref{tab:google_branch} and Table \ref{tab:google_bug} show similar results on branch coverage and bugs. %It is interesting to see that most optimized versions perform worse than the original AFL. 

From Table \ref{tab:google_path}, it is interesting to see that compared with those optimized fuzzers based on AFL (AFLFast, FairFuzz, Radamsa and QSYM), original AFL performs the best on 14 applications in parallel mode with 4 CPU cores. For the total number of paths executed, AFL performs the best and AFLFast performs the worst in parallel mode. While in single mode with one CPU core used, the situation is exactly the opposite, and the original AFL only performs the best on 5 applications, as presented in Table \ref{tab:single_path} of the appendix.

The reason for performance degradation of these optimizations in parallel mode is that their studies lack the consideration for synchronizing the  additional guiding information.
Take AFLFast for example, it models coverage-based fuzzing as Markov Chain, and the times of random mutation for each seed will be computed by a power scheduler. This strategy works well in single mode, %as their evaluation presented, 
but it would fail in parallel mode because the statistics of each fuzzer's scheduler are limited in current thread. %This is critical, because fuzzers generally work in parallel on practical industry applications. 
Our evaluation demonstrates that many optimized fuzzing strategies could be useful in single mode, but fail in the parallel mode even if this is the mode widely used in industry practice. This experiment has been missing by many prior literature studies. A potential solution for this degradation is to synchronize the additional guiding information in their implementation, similar to the work presented in PAFL\cite{liang2018pafl}. 
%when they conclude that the optimization performs better than the traditional AFL

%\vspace{-0.5cm}
\NewEnviron{mytable_google2}[2]{
  \begin{table}[!htbp]
    \caption{#1}
    \scalebox{0.74}[0.74]{%
      \label{tab:#2}
      \begin{tabular}{l|CCCBBCB}
        \toprule
        {\mysize Project}
        & {\mysize AFL}
        & {\mysize AFLFast}
        & {\mysize FairFuzz}
        & {\mysize LibFuzzer}
        & {\mysize Radamsa}
        & {\mysize QSYM}
        & {\mysize \toolThree}\\
        \midrule
        \BODY
        \bottomrule
      \end{tabular}
    }
  \end{table}%
}

\vspace{-0.59cm}
\begin{mytable_google2}{Average number of branches covered by each tool on n Google's fuzzer-test-suite for ten times.}{google_branch}
boringssl     & 3834   & 3635   & 3894   & 3863   & 3880   & 3680   & \textbf{4108 }  \\
c-ares        & \textbf{285  }  & 276    & \textbf{285  }    & 202    & \textbf{285  }    & \textbf{285  }    & \textbf{285  }  \\
guetzli       & 3022   & 2723   & 1514   & \textbf{4016}   & 3177   & 3011   & 3644   \\
lcms          & 3985   & 3681   & 3642   & 3015   & 2857   & 3731   & \textbf{4169 }  \\
libarchive    & 10580  & 9267   & 8646   & 8635   & 11415  & 9416   & \textbf{13949}  \\
libssh        & 614    & 614    & 614    & 573    & 614    & \textbf{636}    & 614    \\
libxml2       & 15204  & 14845  & 14298  & 13346  & 19865  & 14747  & \textbf{21899}  \\
openssl-1.0.1 & 4011   & 3967   & 3996   & 3715   & 4117   & 4032   & \textbf{4673 }  \\
openssl-1.0.2 & 4079   & 4004   & 4021   & 3923   & 4074   & 3892   & \textbf{4216 }  \\
openssl-1.1.0 & 9125   & 9075   & 9123   & 8712   & 9017   & 9058   & \textbf{9827 }  \\
pcre2         & 50558  & 48004  & 49430  & 36539  & 51881  & 36208  & \textbf{53912}  \\
proj4         & 267    & 267    & 267    & 798    & 267    & 261    & \textbf{907  }  \\
re2           & 17918  & 17069  & 17360  & 16001  & 17312  & 16323  & \textbf{19688}  \\
woff2         & 120    & 120    & 120    & 2785   & 120    & 121    & \textbf{3945 }  \\
freetype2     & 53339  & 52404  & 56653  & 57325  & 52715  & 48547  & \textbf{58192}  \\
harfbuzz      & 38163  & 36313  & 43077  & 39712  & 37959  & 38194  & \textbf{44708}  \\
json          & 7048   & 6622   & 5138   & 6583   & 7231   & 7169   & \textbf{7339 }  \\
libjpeg       & 12345  & 11350  & 15688  & 10342  & 12009  & 11468  & \textbf{17071}  \\
libpng        & 4135   & 4393   & 4110   & 4003   & 3961   & 4085   & \textbf{4696 }  \\
llvm          & 55003  & 56619  & 58306  & 57021  & 54312  & 48008  & \textbf{62918}  \\
openthread    & 3109   & 2959   & 2989   & 5421   & 3102   & 3634   & \textbf{5579 }  \\
sqlite        & 2850   & 2847   & 2838   & 3123   & 3012   & 2853   & \textbf{3216 }  \\
vorbis        & 12136  & 13524  & 13053  & 10032  & 11234  & 12849  & \textbf{14318}  \\
wpantund      & 40667  & 40867  & 41404  & 39816  & 40317  & 40556  & \textbf{43217}  \\
\midrule
Total         & 352397 & 345445 & 360466 & 339501 & 354733 & 322764 & \textbf{407090} \\
\midrule  
Improvement &--     & \small{1\% $\downarrow$} & 2\% $\downarrow$ & 3\% $\uparrow$ & 0.6\% $\downarrow$ & 8\%$\downarrow$ & \textbf{16\% $\uparrow$} \\
\end{mytable_google2}

\NewEnviron{mytable_google3}[2]{
  \begin{table}[!htbp]
    \caption{#1}
    \scalebox{0.74}[0.74]{%
      \label{tab:#2}
      \begin{tabular}{l|CBBBCCB}
        \toprule
        {\mysize Project}
        & {\mysize AFL}
        & {\mysize AFLFast}
        & {\mysize FairFuzz}
        & {\mysize LibFuzzer}
        & {\mysize Radamsa}
        & {\mysize QSYM}
        & {\mysize \toolThree}\\
        \midrule
        \BODY
        \bottomrule
      \end{tabular}
    }
  \end{table}%
}

\vspace{-0.5cm}
\begin{mytable_google3}{Average number of unique bugs found by each tool on n Google's fuzzer-test-suite for ten times.}{google_bug}
boringssl     & 0  & 0  & 0  & \textbf{1}  & 0  & 0  & \textbf{1}  \\
c-ares        & \textbf{3}  & 2  & \textbf{3}  & 1  & 2  & 2  & \textbf{3}  \\
guetzli       & 0  & 0  & 0  & \textbf{1}  & 0  & 0  & \textbf{1}  \\
lcms          & 1  & 1  & 1  & \textbf{2}  & 1  & 1  & \textbf{2}  \\
libarchive    & 0  & 0  & 0  & \textbf{1}  & 0  & 0  & \textbf{1}  \\
libssh        & 0  & 0  & 0  & 1  & 0  & 1  & \textbf{2}  \\
libxml2       & 1  & 1  & 1  & \textbf{3}  & 2  & 1  & \textbf{3}  \\
openssl-1.0.1 & 3  & 2  & 3  & 2  & 2  & 3  & \textbf{4}  \\
openssl-1.0.2 & 5  & 4  & 4  & 1  & 5  & 5  & \textbf{6}  \\
openssl-1.1.0 & 5  & 5  & 5  & 3  & 4  & 5  & \textbf{6}  \\
pcre2         & 6  & 4  & 5  & 2  & 5  & 4  & \textbf{8}  \\
proj4         & 2  & 0  & 1  & 1  & 1  & 1  & \textbf{3}  \\
re2           & 1  & 0  & 1  & 1  & 0  & 1  & \textbf{2}  \\
woff2         & 1  & 0  & 0  & \textbf{2}  & 1  & 1  & 1  \\
freetype2     & 0  & 0  & 0  & 0  & 0  & 0  & 0  \\
harfbuzz      & 0  & 0  & \textbf{1}  & \textbf{1}  & 0  & 0  & \textbf{1}  \\
json          & 2  & 1  & 0  & 1  & \textbf{3}  & 2  & \textbf{3}  \\
libjpeg       & 0  & 0  & 0  & 0  & 0  & 0  & 0  \\
libpng        & 0  & 0  & 0  & 0  & 0  & 0  & 0  \\
llvm          & 1  & 1  & \textbf{2}  & \textbf{2}  & 1  & 1  & \textbf{2}  \\
openthread    & 0  & 0  & 0  & \textbf{4}  & 0  & 0  & \textbf{4}  \\
sqlite        & 0  & 0  & 0  & \textbf{3}  & 1  & 1  & \textbf{3}  \\
vorbis        & 3  & \textbf{4}  & 3  & 3  & 3  & \textbf{4}  & \textbf{4}  \\
wpantund      & 0  & 0  & 0  & 0  & 0  & 0  & 0  \\
\midrule  
Total         & 34 & 25 & 30 & 37 & 31 & 33 & \textbf{60} \\
\midrule  
Improvement &--     & \small{26\% $\downarrow$} & 12\% $\uparrow$ & 6\% $\downarrow$ & 9\% $\uparrow$ & 3\%$\downarrow$ & \textbf{76\% $\uparrow$} \\
\end{mytable_google3}

%8. 在多线程场景，radamsa的提升比单线程下要低： （1）radamsa生成的大量无用input会影响afl多线程之间的通信。 （2）afl多线程之间通过共享种子，能够有效自己的性能，这和radamsa所能提供的提升重合了
From the fifth columns of Table \ref{tab:google_path} and Table \ref{tab:single_path}, we find that compared with Radamsa in single mode, the improvement achieved by Radamsa is limited in parallel mode. There are two main reasons:
(1) Too many useless inputs generated by Radamsa slow down the seed-sharing efficiency among all instances of AFL. This seed-sharing mechanism does not exist in single mode.
(2) Some interesting seeds can be created in parallel mode and shared among all instances of AFL. 
These seeds overlap with the inputs generated by Radamsa. So this improvement is limited in parallel mode.

%% 实验结果出来后可能要加一QSYM的分析，也可能不用

For the \toolThree ~which integrates AFL, AFLFast, libFuzzer and Radamsa as base fuzzers and, compared with AFL, AFLFast, FairFuzz, QSYM, LibFuzzer and Radamsa, it shows the strongest robustness and always performs the best.
In total, it discovers 76.4\%, 140\%, 100\%, 81.8\%, 66.7\% and 93.5\% more unique bugs, executes 42.4\%, 61.2\%, 45.8\%, 66.4\%, 29.5\% and 44.2\% more paths and covers 15.5\%, 17.8\%, 12.9\%, 26.1\%, 19.9\% and 14.8\% more branches respectively.
These statistics demonstrate that it helps mitigate performance variation and improves robustness and performance by the ensemble  approach  with  globally  asynchronous  and  locally  synchronous seed synchronization mechanism.

  \subsection{Effects of Different Fuzzing Integration} 
    To study the effects of the globally asynchronous and locally synchronous based seed synchronization mechanism, we conduct a comparative experiment on \toolFour and \toolOne, both ensemble the same base fuzzers (2 instances of AFL, 1 instance of AFLFast, 1 instance of FairFuzz)  in parallel mode with four CPU cores.
To study the effects of different base fuzzers on ensemble fuzzing, we also run \toolFive, \toolTwo ~and \toolThree ~on Google's fuzzer-test-suite for 24 hours 10 times.
To identify unique bugs, we used stack backtraces to deduplicate crashes.
The results are presented in Tables \ref{tab:effect_path}, \ref{tab:effect_branch} and \ref{tab:effect_bug}, which shows the average number of paths executed, branches covered and unique bugs detected by \toolFour, \toolOne, \toolFive, \toolTwo, and \toolThree, ~respectively.
 
Compared with \toolOne, \toolFour ~which ensembles the same base fuzzers — AFL, AFLFast and FairFuzz, but does not implement the seed synchronization mechanism.
\toolFour ~performs much worse on all applications. 
In total, it only executes 68.5\% paths, covers 78.3\% branches and detects 32.4\% unique bugs of EnFuzz-A. 
These statistics demonstrate that the globally asynchronous and locally synchronous based seed synchronization mechanism is critical to the ensemble fuzzing. 

For \toolOne, ~which ensembles AFL, AFLFast and FairFuzz as base fuzzers and implements the seed synchronization with global coverage map, compared with AFL, AFLFast and FairFuzz running in parallel mode with four CPU cores used (as shown in Table \ref{tab:google_path}, Table \ref{tab:google_branch} and Table \ref{tab:google_bug}), it always executes more paths and covers more branches on all applications. In total, it covers 11.3\%, 25.9\% and 13.9\% more paths, achieves 7.2\%, 9.3\% and 4.8\% more covered branches, 
%Benefiting from the increased coverage, \toolOne ~
and triggers 8.8\%, 48\% and 23\% more unique bugs. It reveals that the robustness and performance can be improved even when the diversity of base fuzzers is small.

For the \toolFive ~which integrates AFL, AFLFast, FairFuzz and QYSM as base fuzzers, the results are shown in the fourth columns of Tables \ref{tab:effect_path},  \ref{tab:effect_branch} and  \ref{tab:effect_bug}.
%As mentioned in section \ref{diversity}, the diversity among these base fuzzers is a slightly larger than \toolOne. 
Compared with \toolOne, \toolFive ~covers 1.1\% more paths, executes 1.0\% more branches and triggers 10.8\% more unique bugs than \toolOne. The improvement is significantly smaller on Google's fuzzer-test-suite than on LAVA-M.

\newcolumntype{C}{R{1.4 cm}}
\NewEnviron{mytable_effect}[2]{
  \begin{table}[!htbp]
    \caption{#1}
    \scalebox{0.75}[0.75]{%
      \label{tab:#2}
      \begin{tabular}{l|CCCCC}
        \toprule
        {\mysize Project}
        & {\mysize \toolFour}
        & {\mysize \toolOne}
        & {\mysize \toolFive}
        & {\mysize \toolTwo}
        & {\mysize \toolThree}\\
        \midrule
        \BODY
        \bottomrule
      \end{tabular}
    }
  \end{table}%
}

\vspace{-0.5cm}
\begin{mytable_effect}{Average number of paths covered by each Enfuzz on Google's fuzzer-test-suite for ten times.}{effect_path}
boringssl     &          2590  &          4058  &          3927  &         6782  & \textbf{7136 } \\
c-ares        &          149   &          167   &          159   &         251   & \textbf{253  } \\
guetzli       &          2066  &          3501  &          3472  &         4314  & \textbf{4508 } \\
lcms          &          1056  &          1846  &          1871  &         2253  & \textbf{2433 } \\
libarchive    &          4823  &          14563 &          14501 &         28531 & \textbf{31778} \\
libssh        &          109   &          140   &          152   & \textbf{377}  & \textbf{377  } \\
libxml2       &          11412 &          19928 &          18738 &         33940 & \textbf{35983} \\
openssl-1.0.1 &          3496  &          4015  &          4095  &         4417  & \textbf{4552 } \\
openssl-1.0.2 &          3949  &          4976  & \textbf{5012}  &         4983  & 4991  \\
openssl-1.1.0 &          3850  &          4291  &          4383  &         4733  & \textbf{4801 } \\
pcre2         &          57721 &          81830 &          82642 &         84681 & \textbf{85386} \\
proj4         &          362   &          393   &          399   &         708   & \textbf{709  } \\
re2           &          9053  &          13019 &          14453 &         17056 & \textbf{17155} \\
woff2         &          19    &          25    &          24    &         1314  & \textbf{1324 } \\
freetype2     &          17692 &          22512 &          20134 &         26421 & \textbf{27812} \\
harfbuzz      &          10438 &          14997 &          15019 &         16328 & \textbf{16894} \\
json          &          648   &          1101  &          1183  &         1271  & \textbf{1298 } \\
libjpeg       &          1395  &          2501  &          2475  &         2588  & \textbf{2638 } \\
libpng        &          480   &          601   &          652   &         706   & \textbf{781  } \\
llvm          &          7953  &          9706  &          9668  &         10883 & \textbf{10935} \\
openthread    &          197   &          281   &          743   &         1489  & \textbf{1506 } \\
sqlite        &          279   &          311   &          325   &         598   & \textbf{636  } \\
vorbis        &          928   &          1604  &          1639  &         1673  & \textbf{1699 } \\
wpantund      &          4521  &          5718  &          5731  &         5797  & \textbf{5823 } \\
\midrule  
Total         & 145186 & 212084 & 211397 & 262094 & \textbf{271408} \\
\midrule  
Improvement &--     & \small{46\% $\uparrow$} & 48\% $\uparrow$ & 80\% $\uparrow$ & \textbf{87\% $\uparrow$}  \\
\end{mytable_effect}

%\vspace{-1cm}
\begin{mytable_effect}{Average number of branches covered by each Enfuzz on Google's fuzzer-test-suite for ten times.}{effect_branch}
boringssl     &         3210   &         3996   &         4013   &         4016   & \textbf{4108 }  \\
c-ares        & \textbf{285}   & \textbf{285}   & \textbf{285}   & \textbf{285}   & \textbf{285  }  \\
guetzli       &         2074   &         3316   &         3246   &         3531   & \textbf{3644 }  \\
lcms          &         2872   &         4054   &         4152   &         4098   & \textbf{4169 }  \\
libarchive    &         6092   &         12689  &         11793  &         13267  & \textbf{13949}  \\
libssh        &         613    &         614    & \textbf{640}   &         614    & 614             \\
libxml2       &         14428  &         17657  &         16932  &         21664  & \textbf{21899}  \\
openssl-1.0.1 &         3612   &         4194   &         4204   &         4538   & \textbf{4673 }  \\
openssl-1.0.2 &         4037   &         4176   & \textbf{4292}  &         4202   &         4216    \\
openssl-1.1.0 &         8642   &         9371   &         9401   &         9680   & \textbf{9827 }  \\
pcre2         &         32471  &         51801  &         52751  &         52267  & \textbf{53912}  \\
proj4         &         267    &         267    &         267    & \textbf{907}   & \textbf{907  }  \\
re2           &         16300  &         18070  &         18376  &         19323  & \textbf{19688}  \\
woff2         &         120    &         120    &         121    &         3939   & \textbf{3945 }  \\
freetype2     &         49927  &         55952  &         54193  &         58018  & \textbf{58192}  \\
harfbuzz      &         33915  &         43301  &         43379  &         44419  & \textbf{44708}  \\
json          &         4918   &         7109   &         7146   &         7268   & \textbf{7339 }  \\
libjpeg       &         9826   &         15997  &         15387  &         16984  & \textbf{17071}  \\
libpng        &         3816   &         4487   &         4502   &         4589   & \textbf{4696 }  \\
llvm          &         49186  &         58681  &         58329  &         60104  & \textbf{62918}  \\
openthread    &         2739   &         3221   &         4015   &         5503   & \textbf{5579 }  \\
sqlite        &         2318   &         2898   &         2971   &         3189   & \textbf{3216 }  \\
vorbis        &         10328  &         13872  &         13993  &         14210  & \textbf{14318}  \\
wpantund      &         33749  &         41537  &         41663  &         43104  & \textbf{43217}  \\
\midrule  
Total         & 295745 & 377665 & 376051 & 399719 & \textbf{407090} \\
\midrule  
Improvement &--     & \small{27\% $\uparrow$} & 28\% $\uparrow$ & 35\% $\uparrow$ & \textbf{38\%} $\uparrow$  \\
\end{mytable_effect}

%\vspace{-0.4cm}
\begin{mytable_effect}{Average number of bugs found by each Enfuzz on Google's fuzzer-test-suite for ten times.}{effect_bug}
boringssl     &          0  &          0  &          0  & \textbf{1}  & \textbf{1}  \\
c-ares        &          1  & \textbf{3}  &          2  & \textbf{3}  & \textbf{3}  \\
guetzli       &          0  &          0  & \textbf{1}  & \textbf{1}  & \textbf{1}  \\
lcms          &          0  &          1  &          1  & \textbf{2}  & \textbf{2}  \\
libarchive    &          0  &          0  & \textbf{1}  & \textbf{1}  & \textbf{1}  \\
libssh        &          0  &          0  & \textbf{2}  & \textbf{2}  & \textbf{2}  \\
libxml2       &          1  &          1  &          1  &          2  & \textbf{3}  \\
openssl-1.0.1 &          0  &          3  &          3  & \textbf{4}  & \textbf{4}  \\
openssl-1.0.2 &          3  &          5  &          5  &          5  & \textbf{6}  \\
openssl-1.1.0 &          2  &          5  &          5  & \textbf{6}  & \textbf{6}  \\
pcre2         &          3  &          6  &          6  &          7  & \textbf{8}  \\
proj4         &          0  &          2  &          2  &          2  & \textbf{3}  \\
re2           &          0  &          1  &          1  &          2  & \textbf{2}  \\
woff2         &          0  & \textbf{1}  & \textbf{1}  & \textbf{1}  & \textbf{1}  \\
freetype2     &          0  &          0  &          0  &          0  & \textbf{0}  \\
harfbuzz      &          0  & \textbf{1}  & \textbf{1}  & \textbf{1}  & \textbf{1}  \\
json          &          1  &          2  &          2  &          2  & \textbf{3}  \\
libjpeg       &          0  &          0  &          0  &          0  & \textbf{0}  \\
libpng        &          0  &          0  &          0  &          0  & \textbf{0}  \\
llvm          &          0  &          1  &          1  & \textbf{2}  & \textbf{2}  \\
openthread    &          0  &          0  &          1  &          3  & \textbf{4}  \\
sqlite        &          0  &          1  &          1  &          2  & \textbf{3}  \\
vorbis        &          1  & \textbf{4}  & \textbf{4}  & \textbf{4}  & \textbf{4}  \\
wpantund      & 0  & 0  & 0  & 0  & 0  \\
\midrule  
Total         & 12 & 37 & 41 & 53 & \textbf{60}  \\
\midrule  
Improvement &--     & \small{208\% $\uparrow$} & 242\% $\uparrow$ & 342\% $\uparrow$ & \textbf{400\% $\uparrow$}  \\
\end{mytable_effect}

The reason for performance degradation between experiments on LAVA-M and Google fuzzer-test-suite is that the base codes of the four applications (who, uniq, base64 and md5sum) in LAVA-M are small (2K-4K LOCs). The concolic execution engine works well on them, but usually performs the opposite or even hangs on real projects in fuzzer-test-suite whose code base easily reaches 100k LOCs.

For the \toolTwo ~which integrates AFL, AFLFast, FairFuzz and libFuzzer as base fuzzers, the results are presented in the seventh columns of Tables \ref{tab:effect_path},  \ref{tab:effect_branch} and  \ref{tab:effect_bug}.
As mentioned in section \ref{diversity}, the diversity among these base fuzzers is much larger than with \toolOne. 
Compared with \toolOne, \toolTwo ~always performs better on all target applications. In total, it covers 23.6\% more paths, executes 5.8\% more branches and triggers 42.4\% more unique bugs than \toolOne.

For the \toolThree ~which integrates AFL, AFLFast, libFuzzer and Radamsa as base fuzzers, the diversity is the largest because they cover all three diversity heuristics. Compared with \toolTwo, it performs better and covers 3.6\% more paths, executes 1.8\% more branches and triggers 13.2\% more unique bugs. Both \toolThree ~and \toolTwo ~performs better than \toolFive.  These statistics demonstrate that the more diversity among these base fuzzers, the better the ensemble fuzzer should perform. For real applications with a large code base, compared with hybrid conclic fuzzing or ensemble fuzzing with symbolic execution, the ensemble fuzzing without symbolic execution may perform better.
    \label{effect}
 % \subsection{Real-world Applications CVE Mining} 
 \subsection{Fuzzing Real-World Applications} 
    We apply \toolname to fuzz more real-world applications from GitHub and commercial products from Cisco, some of which are well-fuzzed projects such as the image processing library libpng and libjepg, the video processing library libwav, the IoT device communication protocol libiec61850 used in hundreds of thousands of cameras, etc. 
\toolname also performs well. Within 24 hours, besides the coverage improvements, \toolname finds \bugnum more unknown real bugs including \cvenum successfully registered as CVEs, as shown in Table \ref{tab:CVE}. All of these new bugs and security
vulnerabilities are detected in a 64-bit machine with 36 cores (Intel(R) Xeon(R)  CPU E5-2630 v3@2.40GHz), 128GB of main memory, and Ubuntu 16.04 as the host OS.

\newcolumntype{B}{>{\arraybackslash}p{1.0cm}}
\newcolumntype{C}{>{\arraybackslash}p{0.75cm}}
\NewEnviron{mytable_cve}[2]{
  \begin{table}[!htbp]
    \caption{#1}
    \scalebox{0.8}[0.9]{%
      \label{tab:#2}
      \begin{tabular}{l|CCCBCC}
        \toprule
        {\mysize Project}
        & {\mysize AFL}
        & {\mysize AFLFast}
        & {\mysize FairFuzz}
        & {\mysize LibFuzzer}
        & {\mysize QSYM}
        & {\mysize \toolThree}\\
        \midrule
        \BODY
        \bottomrule
      \end{tabular}
    }
  \end{table}%
}

\begin{mytable_cve}{Unique previously unknown bugs detected by each tool within 24 hours on some real-world applications.}{real_bug}
Bento4\_mp4com &          5  &          4  &          5  &          5  &          4  & \textbf{6}  \\
Bento4\_mp4tag     &          5  &          4  &          4  &          5  &          4  & \textbf{7}  \\
bitmap             &          1  &          1  &          1  &          0  &          1  & \textbf{2}  \\
cmft               &          1  &          1  &          0  &          1  &          0  & \textbf{2}  \\
ffjpeg             &          1  &          1  &          1  &          0  &          1  & \textbf{2}  \\
flif               &          1  &          1  &          1  &          2  &          1  & \textbf{3}  \\
imageworsener      & \textbf{1}  &          0  &          0  &          0  & \textbf{1}  & \textbf{1}  \\
libjpeg-05-2018    &          3  &          3  &          3  &          4  &          3  & \textbf{5}  \\
libiec61850        &          3  &          2  &          2  &          1  &          2  & \textbf{4}  \\
libpng-1.6.34      &          2  &          1  &          1  &          1  &          2  & \textbf{3}  \\
libwav\_wavgain    &          3  &          2  &          3  &          0  &          2  & \textbf{5}  \\
libwav\_wavinfo    &          2  &          1  &          2  &          4  &          2  & \textbf{5}  \\
LuPng              &          1  &          1  &          1  &          3  &          1  & \textbf{4}  \\
pbc                &          5  &          5  &          6  &          7  &          6  & \textbf{9}  \\
pngwriter          &          1  &          1  &          1  &          1  & \textbf{2}  & \textbf{2}  \\
\midrule  
total              & 35 & 28 & 31 & 34 & 32 & \textbf{60} \\
\end{mytable_cve}

As a comparison, we also run each tool on those real-world applications to detect unknown vulnerabilities. The results are presented in table \ref{tab:real_bug}. \toolThree ~found all 60 unique bugs, while other tools only found a portion of these bugs. Compared with AFL, AFLFast, FairFuzz, LibFuzzer and QSYM, \toolThree ~detected 71.4\%, 114\%, 93.5\%, 76.4\%, 87.5\% more unique bugs respectively.
The results demonstrate the effectiveness of \toolThree ~in detecting real vulnerabilities in more general projects. For example, in the well-fuzzed projects libwav and libpng, we can still detect 13 more real bugs, 7 of which are assigned as CVEs. We give an analysis of the project libpng for a more detailed illustration. 
libpng is a widely used C library for reading and writing PNG image files. It has been fuzzed many times and is one of the projects in Google's OSS-Fuzz, which means it has been continually fuzzed by multiple fuzzers many times. But with \toolThree, we detect three vulnerabilities, including one segmentation fault, one stack-buffer-overflow and one memory leak. The first two vulnerabilities were assigned as CVEs (CVE-2018-14047, CVE-2018-14550).

In particular, CVE-2018-14047 allows remote attackers to cause a segmentation fault via a crafted input. We analyze the vulnerability with AddressSanitizer and find it is a typical memory access violation. 
%As Listing \ref{gdb} shows, 
The problem is that in function \texttt{png\_free\_data} in line 564 of png.c, the info\_ptr attempts to access an invalid area of memory.
The error occurs in \texttt{png\_free\_data} during the free of text-related data with specifically crafted files, and causes reading of invalid or unknown memory, as show in Listing \ref{libpng}.
The new vulnerabilities and CVEs in the IoT device communication protocol libiec6185 can also crash the service and have already been confirmed and  repaired.

\begin{figure}[!htbp]
\begin{lstlisting}[language = C, numbers = none,
        commentstyle=\color{red!50!green!50!blue!50},frame=shadowbox,
        rulesepcolor=\color{red!20!green!20!blue!20},basicstyle=\ttfamily,breaklines=true,
        extendedchars=true,caption={The error code of libpng for CVE-2018-14047 },label={libpng}]
#ifdef PNG_TEXT_SUPPORTED
/* Free text item num or (if num == -1) all text items */
   	if (info_ptr->text != NULL &&
       	((mask & PNG_FREE_TEXT) & info_ptr->free_me) != 0)
\end{lstlisting} 
\end{figure}

\begin{figure}[!htbp]
\centering
\begin{minipage}[!htbp]{0.40\textwidth}
     \centering
     \includegraphics[width=1.0\textwidth]{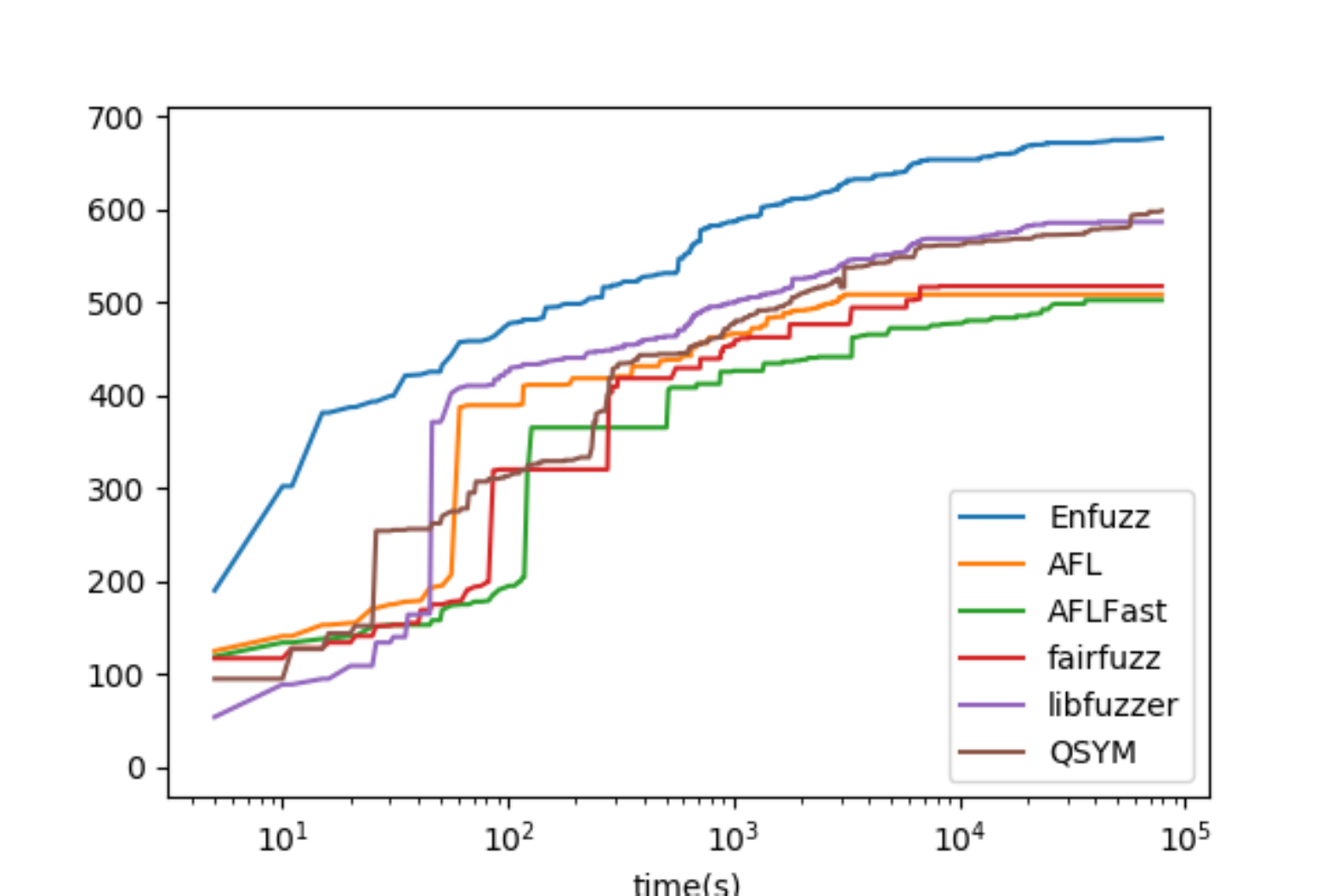}
     \small{(a) Number of paths over time}
\end{minipage}
\begin{minipage}[!htbp]{0.40\textwidth}
     \centering
     \includegraphics[width=1.0\textwidth]{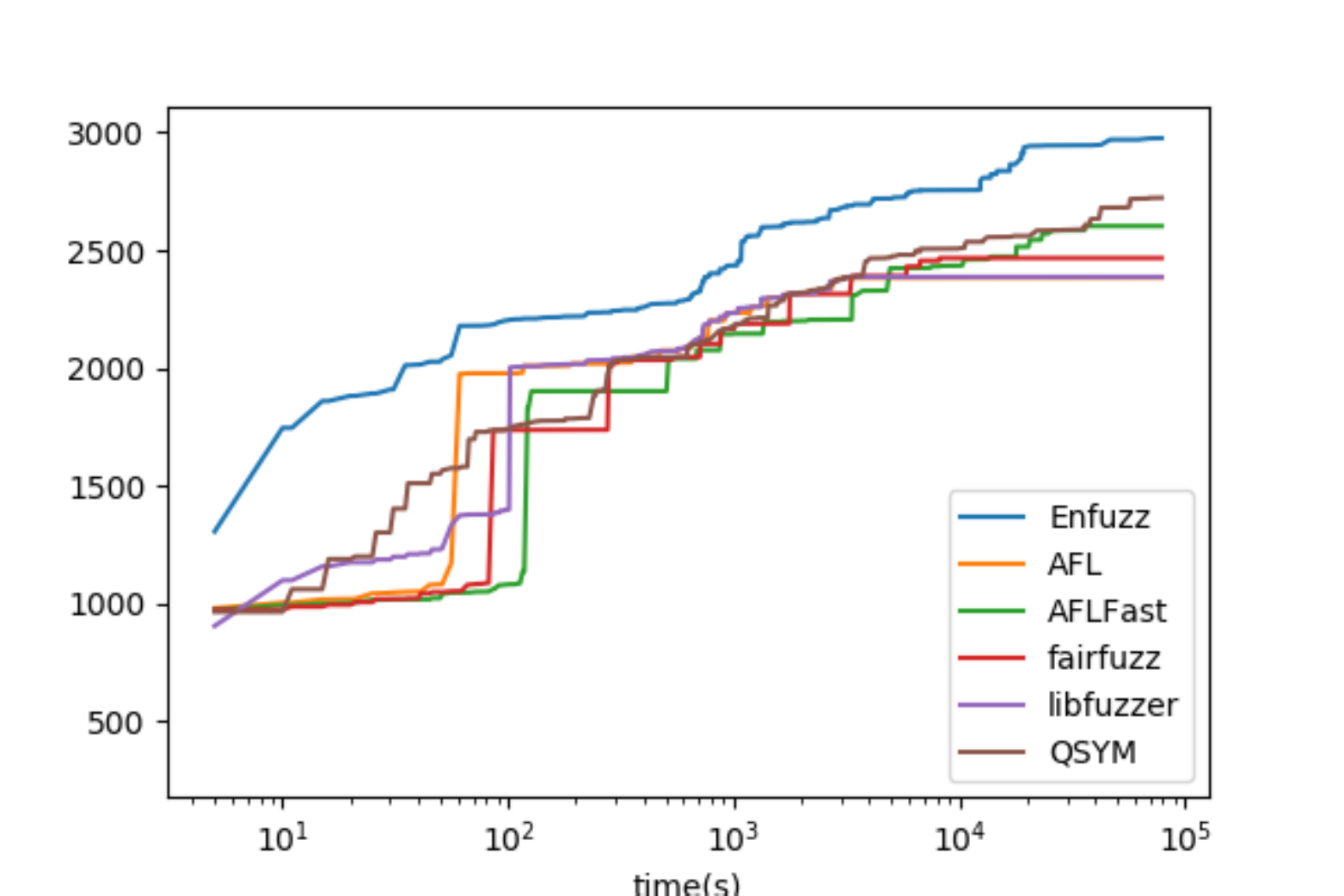}     
     \small{(b) Number of branches over time}
\end{minipage}
\caption{Performance of each fuzzer over time in libpng. Each fuzzer runs in four CPU cores for 24 hours.}
\label{fig:performance in libpng}
\end{figure}

We also apply each base fuzzer (AFL, AFLFast, FairFuzz, libFuzzer and QSYM) to fuzz libpng separately, the above vulnerability is not detected.
%, as is consistent with previous studies.
To trigger this bug, 6 function calls and 11 compares (2 for integer, 1 for boolean and 8 for pointer) are required. It is difficult for other fuzzers to detect bugs in such deep paths without the seeds synchronization of EnFuzz.
The performances of these fuzzers over time in libpng are presented in Figure \ref{fig:performance in libpng}. The results demonstrate that generalization and scalability limitations exist in these base fuzzers -- the two optimized fuzzers AFLFast and FairFuzz perform worse than the original AFL for libpng, while \toolThree ~performs the best. Furthermore, except for those evaluations on benchmarks and real projects, \EnFuzz~ had already been deployed in industry practice, and more new CVEs were being continuously reported.

\newcolumntype{C}{>{\arraybackslash}p{0.7 cm}}
\NewEnviron{CVETable}[2]{
  \begin{table}[!htp]
    \caption{#1}
    \scalebox{0.95}[0.9]{%
      \label{tab:#2}
      \begin{tabular}{l|C|p{4.5 cm}}
        \toprule
        {\mysize Project}
        & {\mysize Count}
        & {\mysize CVE-2018-Number}\\
        \midrule
        \BODY
        \bottomrule
      \end{tabular}
    }
  \end{table}%
}

\begin{CVETable}{The \cvenum CVEs detected by \toolname in 24 hours.}{CVE}
Bento4\_mp4com   & 6		&14584, 14585, 14586, 14587, 14588, 14589                                        \\
Bento4\_mp4tag   & 6		&13846, 13847, 13848, 14590, 14531, 14532                                        \\
bitmap           & 1        &17073	                                                                            \\
cmft		     & 1		&13833	                                                                            \\           
ffjpeg           &1         &16781	                                                                            \\
flif	         & 1	    &12109	                                                                            \\
imageworsener    &1         &16782	                                                                            \\
libjpeg-05-2018  & 4        &11212, 11213, 11214, 11813                                                      \\
libiec61850      & 3        &18834, 18937, 19093                                                             \\
libpng-1.6.34    & 2		&14048, 14550	                                                                \\
libwav\_wavgain	 & 2		&14052, 14549	                                                                  \\
libwav\_wavinfo	 & 3		&14049, 14050, 14051	                                                              \\
LuPng	         & 3 	    &18581, 18582, 18583	                                                              \\
pbc			     & 9		&14736, 14737, 14738, 14739, 14740, 14741, 14742, 14743, 14744                    \\
pngwriter	     & 1		&14047\\
\end{CVETable}

%    \subsection{Does performance variation Exist? }
%        \input{part-eval_generalization}
%    \subsection{How does diversity affect the performance of fuzzing?} %\vspace{-.02cm}
%    	\label{diversity}
%    	\input{part-eval_diversity}
%    \subsection{Can Ensemble Fuzzer Perform Better?} %\vspace{-.02cm}
%    	\input{part-eval_enfuzz}

\section{Discussion}
\label{Discussion}
  
%% base fuzzers 的差异性 和 ensemble method 是集成fuzzer的核心， 这里我们只给出了几种差异性的定义，未来有更多的研究点。 此外， 集成的方法也 过于简单， 研究的点也很多
Based on benchmarks such as LAVA-M and Google's fuzzer-test-suite, and several real projects, we demonstrate that this ensemble fuzzing approach outperforms any base fuzzers. However, some limitations still threaten the performance of ensemble fuzzing. The representative limitations and the workarounds are discussed below.

The first potential threat is the insufficient and imprecise diversity of base fuzzers. Section \ref{Base Fuzzer Selection} describes our base fuzzer selection, we propose three different heuristics  to indicate  diversity of base fuzzers, including diversity of coverage information granularity, diversity of input genera-tion strategy, and diversity of seed mutation selection strategy.
According to these three heuristics, we select AFL, AFLFast, FairFuzz, libFuzzer, Radamsa and QSYM as the base fuzzers. %They are open-source and demonstrate the issue of performance variation.
Furthermore, we implement four prototypes of ensemble fuzzing and demonstrate that the greater the diversity of base fuzzers, the better the ensemble fuzzer performs. However, these three different heuristics of diversity may be insufficient. More diversity measures need to be proposed in future work. For example, initial seeds determine the initial direction of fuzzing and, thus, are significantly important for fuzzing, especially for mutation-based fuzzers. Some fuzzers utilize initial seeds generated by symbolic execution \cite{wang2018safl, ognawala2018improving} while some other fuzzers utilize initial seeds constructed by domain experts or grammar specifications. However, we select base fuzzers manually according to the initial diversity heuristic, which is also not accurate enough. 
%the quantification of diversity value is based on the benchmark for preliminary reference, which is also not accurate enough. 
%It could be improved by accumulated evaluation of more projects and integration of variation results on more types of metrics. With fuzzing statics on more projects and more dimensions, we can get a more accurate diversity quantification based on the evaluation formula 1 and 2 described in the section 4.1.  %Different initial seed generation strategies may have critical influences on the diversity of fuzzers. In ensemble learning, a number of diversity measures have been designed,
%we could also get inspiration from them and propose more diversity measures.

A possible solution to this threat is to quantify the initial diversity value among different fuzzers for more accurate selection. 
As defined in \cite{benjamin2014probability}, the variance or diversity is a measure of the distance of the data in relation to the average. The average standard deviation of a data set is a percentage that indicates how much, on average, each measurement differs from the other. 
%Take path coverage for example, as shown in Table \ref{tab:single_path}, 
To evaluate the diversity of different base fuzzers, we can choose the most widely used AFL and its path coverage as a baseline and then calculate standard deviation of each tool from this baseline on the Google fuzzing-test-suite. Then we can calculate the standard deviation of these values as the initial measure of diversity for each base fuzzer, as presented in formula (\ref{eq:diversity}) and (\ref{eq:mean}),  
where \(n\) means the number of applications fuzzed by these base fuzzers, \(p_i\) means the number of paths covered by the current fuzzer of the target application \(i\) and \(p_{A_i}\) means the number of paths covered by AFL of the application \(i\).

\begin{equation}\label{eq:mean}
mean = \frac{1}{n} \sum_{i=1}^{n}{ \frac{p_i - p_{A_i}}{p_{A_i}} }
\end{equation}

\begin{equation}\label{eq:diversity}
diversity = \frac{1}{n} \sum_{i=1}^{n}{ {( \frac{p_i - p_{A_i}}{p_{A_i}} - mean )} ^{ 2 }  }
\end{equation}

Take the diversity of AFLFast, FairFuzz, Radamsa, QSYM, and libFuzzer for example, as shown in the statistics presented in Table \ref{tab:single_path} of the appendix, compared with AFL on different applications, 
the diversity of AFLFast is 0.040; 
the diversity of FairFuzz is 0.062; 
the diversity of Radamsa is 0.197; 
the diversity of QSYM is 0.271; 
the diversity of libFuzzer is 11.929. 
In the same way, the deviation on branches covered and the bugs detected can be calculated. We can add these three values together with different weight for the final diversity quantification. For example, the bug deviation should be assigned with more weights, because from prior research, coverage metrics (the number of paths or branches) are not necessarily correlated well with bugs found.  A more advanced way to evaluate the amount of diversity would be to count how many paths/branches/bugs were found by one fuzzer and not by any of the others.

The second potential threat is the mechanism scalability of the ensemble architecture. 
Section \ref{Ensemble Architecture Design} describes the ensemble architecture design, and proposes the globally asynchronous and locally synchronous based seed synchronization mechanism.
The seed synchronization mechanism focuses on enhancing cooperation among these base fuzzers during their fuzzing processes.
With the help of seeds sharing, the performance of ensemble fuzzing is much improved and is better than any of the constituent base fuzzers with the same computing resources usage.
However, this mechanism can still be improved for better scalability on different applications and fuzzing tasks. \EnFuzz ~only synchronizes the coarse-grained information -- interesting seeds, rather than the fine-grained information.
For example, we could synchronize the execution trace and array index values of each base fuzzer to improve their effectiveness in cooperation. 
Furthermore, we currently select and mix base fuzzers manually according to three heuristics. When scaled to arbitrary number of cores, it should be carefully investigated with huge number of empirical evaluations. A possible solution is that the base fuzzers will be dynamically selected and initiated with different number of cores according to the real-time number of paths/branches/bugs found individually by each fuzzer. In the beginning, we have a set of different base fuzzers; then Enfuzz selects n (this number can be configured) base fuzzers randomly. If one fuzzer cannot contribute to coverage for a long time, then it will be terminated, and one new base fuzzer from the sets will be setup for fuzzing or the existing live base fuzzer with better coverage will be allocated with more CPU cores. 

%The base fuzzers should be selected dynamically according to the real-time number of paths/branches/bugs found individually by each fuzzer. For example, if one fuzzer can not contribute to coverage for a long time, then it should be terminated, and one new base fuzzers should be setup for fuzzing.

We can also apply some effective ensemble mechanisms in ensemble learning such as Boosting to ensemble fuzzing to improve the scalability.
Boosting is a widely used ensemble mechanism which will reweigh the base learner dynamically to improve the performance of the ensemble learner: examples that are misclassified gain weight and examples that are classified correctly lose weight.
To implement this idea in ensemble fuzzing, we could start up a \textit{master} thread to monitor the execution statuses of all base fuzzers and record more precise information of each base fuzzer, then reassign each base fuzzer some interesting seeds accordingly.

%% 根据review1 关于集成参数的意见，加了以下两段说明（不知道要不要加）
For the number of base fuzzers and parameters in ensemble fuzzing implementation, it is scalable for integration of most fuzzers. Theoretically, the more base fuzzers with diversity, the better ensemble fuzzing performs. We only use four base fuzzers in our evaluation with four CPU cores. The more computing resources we get, higher performance the fuzzing practice acquires. %The influence of the number of base fuzzers can be explored in the future. 
Furthermore, in our implementation, 
we have tried different values of period time, and the results are very sensitive to the specific setting of this value. It only affects the performance in the beginning, but affects little in the end.
%the time period of synchronously seed sharing is 120 seconds. 
%This parameter is closely related to the size of the target applications. A shorter period consumes too many resources, which leads to a decrease in fuzzing performance. 
%A longer period results in untimely seed synchronization, which also affects fuzzing performance. We can use the global seed pool status sampling of early fuzzing stages for period optimization. 
Furthermore, refering to the GALS system design, we can also allocate a different synchronization frequency for each local fuzzer dynamically. %How to find a more suitable balance point dynamically need to be explored in the future.

\section{Conclusion}
\label{Conclusion}
  In this paper, we systematically investigate the practical ensemble fuzzing strategies and the effectiveness of ensemble fuzzing of various fuzzers. %Ensembling multiple diverse fuzzing strategies with the global asynchronous and local synchronous based seed sharing mechanism solves the issue of performance variation in different base fuzzers, and obtains better performance than that of any constituent base fuzzer alone. 
Applying the idea of ensemble fuzzing, we bridge two gaps.
First, we come up with a method for defining the diversity of base fuzzers and propose a way of selecting  a diverse set of base fuzzers. 
Then, inspired by AFL in parallel mode, we implement a concrete ensemble architecture with one effective ensemble mechanism, a seed synchronization mechanism.
%For evaluation, we implement three prototypes of ensemble fuzzing. 
%We evaluate them with base fuzzers on a third-party benchmark, Google's fuzzer-test-suite, which consists of real-world applications.
\EnFuzz ~always outperforms other popular base fuzzers in terms of unique bugs, path and branch coverage with the same resource usage. 
%We also find that some existing optimizations for fuzzing strategies work well in single mode, but fail in parallel mode.  
%due to lack of information synchronization among threads.  
\EnFuzz ~has found \bugnum new bugs in several well-fuzzed projects and \cvenum new CVEs were assigned.
Our ensemble architecture can be easily utilized to integrate other base fuzzers for industrial practice. 

Our future work will focus on three directions: the first is to try some other heuristics and more accurate accumulated quantification of diversity in base fuzzers; the second is to improve the ensemble architecture with more advanced en- semble mechanism and synchronize more fine-grained information; the last is to improve the ensemble architecture with intelligent resource allocation such as  dynamically adjusting the synchronization period for each base fuzzer, and allocating more CPU cores to the base fuzzer that shares more interesting seeds. %and release the source code for academic study and industry practice in vulnerability detection. 

{\normalsize \bibliographystyle{acm}
\bibliography{ADG}

\appendix
\section{Preliminary demonstration of diversity among base fuzzers}
	\label{diversity}
	To help select base fuzzers with larger diversity, we need to estimate the diversity between each base fuzzer. In general, the more differently they perform on different
applications, the more diversity among these base fuzzers.
Accordingly, we first run each base fuzzer in single mode, with one CPU core on Google's fuzzer-test-suite for 24 hours. 
Table \ref{tab:single_path} and Table \ref{tab:single_branch} show the number of paths and branches covered by AFL, AFLFast, FairFuzz, libFuzzer, Radamsa and QSYM. Table \ref{tab:single_crash} shows the corresponding number of unique bugs. Below we present the performance effects of the three diversity heuristics proposed in Section 4.1 in detail. 

%Then, we choose AFL as baseline, and calculate diversity of each base fuzzers between AFL, as described in formula \ref{eq:mean} and formula \ref{eq:diversity}.
%the diversity between AFL and  AFLFast is 0.040; 
%the diversity between AFL and  FairFuzz is 0.062; 
%the diversity between AFL and combining Radamsa with AFL is 0.197; 
%the diversity of between AFL and QSYM is 0.271; 
%the diversity of between AFL and libFuzzer is 11.929; 

\textit{1) Effects of seed mutation and seed selection strategy -- what kind of mutation and selection strategy you use, what kind of path and branch you would cover}
The first three columns of Table \ref{tab:single_path} show the performance of the AFL family tools. Their differences are the seed mutation and seed selection strategies. The original AFL performs the best on 5 applications, but  performs the worst on other 10 applications. 
%However, according to literature reviews, those optimizations should outperform AFL. 
%More specifically, 
AFLFast performs the best on 13 applications, and only performs the worst on 4 applications. FairFuzz also performs the best on 8 applications, but the worst on the other 9 applications.
Although the total number of paths covered improves slightly, the performance variation on each application is huge, ranging from -57\% to 38\% in single cases.
%, as demonstrated in Figure \ref{fig:paths executed compared with AFL}. 

From the first three columns in Table \ref{tab:single_branch} and Table \ref{tab:single_crash}, we get the same observation that the performance of these optimized fuzzers varies significantly on different applications. Although the total number of covered branches and unique crashes improves slightly, the deviation of each application is huge.
AFLFast selects seeds that exercise low-frequency paths to mutate more times. Take project lcms for example, this seed selection strategy exercises more new paths by avoiding covering ``hot paths'' too many times, but on project libarchive, its ``hot path'' may be the key to further paths.
FairFuzz mutates seeds to hit rare branches. Take project libxml2 for example, the rare branch fuzzing strategy guides FairFuzz into deeper areas and covers more branches. However, on libarchive, this strategy fails. FairFuzz spends much time in deep paths and branches, ignoring breadth search. Unlike libxml2, the breadth first search strategy of other fuzzers is more effective on libarchive. In general, the mutation and selection strategy decides the depth and breath of the covered branch and path.

\textit{2) Effects of coverage information granularity--what kind of guided information you use, what kind of coverage metric you improve.}
%4. 与AFL相比，libfuzzer的path上比它好， branch却不如afl，因为libfuzzer是path-coverage guided的
The diversity between AFL and libFuzzer is their coverage information granularity.
According to the fourth column of Table \ref{tab:single_path}, we find that compared with AFL, libFuzzer performs better on 17 applications, and covers 30.3\% more paths in total. However, according to the fourth column of the Table \ref{tab:single_branch}, compared with AFL, libFuzzer only performs better on 11 applications, which means on 6 applications, libFuzzer covers more paths but less branches. For total branch count, AFL covers 7.3\% more than libFuzzer.
The reason is that AFL mutates seed by tracking edge hit counts while libFuzzer utilizes the SanitizerCoverage instrumentation to track block hit counts. AFL prefers to cover more branches while libFuzzer is better at executing more paths. In general, edge-guided means more branches covered, and block-guided means more paths covered.

\vspace{-0.5cm}
\newcolumntype{C}{R{0.9cm}}
\NewEnviron{mytable_single}[2]{
  \begin{table}[!htbp]
    \caption{#1}
    \scalebox{0.83}[0.83]{%
      \label{tab:#2}
      \begin{tabular}{l|CCCCCC}
        \toprule
        {\mysize Project}
        & {\mysize AFL}
        & {\mysize AFLFast}
        & {\mysize FairFuzz}
        & {\mysize libFuzzer}
        & {\mysize Radamsa}
        & {\mysize QSYM}\\
        \midrule
        \BODY
        \bottomrule
      \end{tabular}
    }
  \end{table}%
}

\begin{mytable_single}{Average number of paths for single mode. }{single_path} 
boringssl     &          1334  &          1674  &          1760  & \textbf{3528}   &          1682  &          1207  \\
c-ares        &          80    &          84    &          88    & \textbf{123}    &          78    &          72    \\
guetzli       &          1382  &          1090  &          1030  & \textbf{1773}   &          1562  &          1268  \\
lcms          &          656   & \textbf{864}   &          434   &          338    &          550   &          605   \\
libarchive    &          3756  &          2834  &          1630  & \textbf{10124}  &          4570  &          3505  \\
libssh        &          64    &          68    &          62    & \textbf{201}    &          63    &          87    \\
libxml2       &          5762  &          7956  &          8028  & \textbf{19663}  &          9392  &          5098  \\
openssl-1.0.1 & \textbf{2397}  &          2103  &          2285  &          1709   &          2303  &          2330  \\
openssl-1.0.2 &          2456  & \textbf{2482}  &          2040  &          1881   &          2108  &          1947  \\
openssl-1.1.0 &          2439  &          2380  & \textbf{2501}  &          1897   &          2311  &          2416  \\
pcre2         &          32310 &          35288 &          36176 &          20981  & \textbf{37850} &          24501 \\
proj4         &          220   &          218   &          218   & \textbf{334}    &          182   &          208   \\
re2           &          5860  &          6014  &          5016  & \textbf{6327}   &          5418  &          5084  \\
woff2         &          14    &          10    &          12    & \textbf{224}    &          10    &          15    \\
freetype2     &          7748  &          10939 &          10714 & \textbf{16360}  &          9825  &          7188  \\
harfbuzz      &          6793  &          8068  &          8668  & \textbf{10800}  &          5688  &          6881  \\
json          &          466   &          412   &          408   &          499    & \textbf{564}   &          504   \\
libjpeg       &          704   & \textbf{979}   &          722   &          448    &          634   &          638   \\
libpng        &          170   &          159   &          76    &          263    &          493   & \textbf{577}   \\
llvm          &          4830  & \textbf{5760}  &          5360  &          5646   &          4593  &          4096  \\
openthread    &          104   &          123   &          127   & \textbf{976}    &          144   &          141   \\
sqlite        &          179   &          193   &          172   & \textbf{431}    &          256   &          180   \\
vorbis        &          891   & \textbf{1122}  &          821   &          848    &          875   &          898   \\
wpantund      &          2959  &          3048  & \textbf{3513}  &          3510   &          3146  &          2975  \\
\midrule  
Total         & 83575 & 93867 & 91862 & \textbf{108884} & 94296 & 72422 \\  
\end{mytable_single}

%\vspace{-0.5cm}
\begin{mytable_single}{Average number of branches for single mode.}{single_branch}
boringssl     &         2645   &          3054   &          3115   &          3608   &  \textbf{3641}  &          2539   \\
c-ares        & \textbf{126}   &          122    &  \textbf{126}   &          100    &  \textbf{126}   &  \textbf{126}   \\
guetzli       &         1913   &          1491   &          1428   &  \textbf{2774}  &          2118   &          1906   \\
lcms          &         2216   & \textbf{2755}   &          935    &          2661   &          1661   &          2075   \\
libarchive    &         4906   &          3961   &          2387   &          3561   &  \textbf{5263}  &          4366   \\
libssh        &         604    &          604    &          604    &          518    &          604    & \textbf{626}    \\
libxml2       &         10082  &          12407  &          12655  &          13037  &  \textbf{14287} &          9779   \\
openssl-1.0.1 &         3809   &          3879   & \textbf{3901}   &          2591   &          2993   &          3829   \\
openssl-1.0.2 &         3978   &          4015   &          3883   &          2308   &  \textbf{4068}  &          3796   \\
openssl-1.1.0 &         8091   &          8132   &          8212   &          7810   &  \textbf{8292}  &          8032   \\
pcre2         &         27308  &          29324  &          28404  &          13463  &  \textbf{30615} &          19557  \\
proj4         &         264    &          260    &          260    & \textbf{683}    &          264    &          258    \\
re2           &         15892  &          15970  &          15073  &          11369  &  \textbf{16485} &         14477  \\
woff2         &         114    &          112    &          114    & \textbf{1003}   &          114    &          115    \\
freetype2     &         36798  &          44028  &          45319  &          45541  &  \textbf{49468} &          33492  \\
harfbuzz      &         16872  &          16051  & \textbf{19045}  &          18659  &          16782  &          16886  \\
json          &         4462   &          3626   & \textbf{4846}   &          4547   &          4821   &          4538   \\
libjpeg       &         6865   &          8495   &          4028   & \textbf{8828}   &          6982   &          6377   \\
libpng        &         1917   &          1878   &          1135   &          1651   &          2126   & \textbf{2294}   \\
llvm          &         54107  &          55697  & \textbf{57356}  &          51548  &          53427  &          47226  \\
openthread    &         2062   &          2473   &          2646   & \textbf{5295}   &          2231   &          2410   \\
sqlite        &         2706   & \textbf{2784}   &          2771   &          2178   &          2190   &          2709   \\
vorbis        &         11836  & \textbf{13561}  &          12605  &          5902   &          11217  &          12531  \\
wpantund      &         36059  &          36620  & \textbf{37269}  &          28694  &          37075  &          35960  \\
\midrule  
Total         & 255631 & 271299 & 268116 & 238329 & \textbf{276850} & 235903 \\  
\end{mytable_single}

\textit{3) Effects of Input generation strategy--what kind of generation strategy you use, what kind of corresponding application you fuzz better.}
%5. 与AFL相比，加入radamsa能够在某些特定的项目上得到有效的提升，以libxml为例，radamsa具备一些领域知识，能够有效的生成大量类似于xml一样的结构化数据，对于complex format input类型，generation-based + mutation-based 能够比较有效。 但在一些非这种场景下，radamsa生成的大量无用input可能会影响afl的效率
The diversity between AFL and Radamsa is the input generation strategy.
From the fifth columns of Table \ref{tab:single_path} and Table \ref{tab:single_branch}, compared with AFL, the plenty of inputs generated by Radamsa have some side effects on most target applications (14 applications). Too many extra inputs will slow down the execution speed of the fuzzer. However, for some applications, the inputs generated by Radamsa will improve the performance effectively. Take libxml2 for example, Radamsa has some domain knowledge that prefers to generate some structured data and specific complex format data. These domain knowledge are not available in most mutation-based fuzzers, and this is a critical disadvantage of AFL. But with the help of generation-based fuzzers, the performance of AFL can be improved greatly. %\textbf{In general, generation-based fuzzers can be better at applications with complex structured inputs with some domain information.}

\begin{mytable_single}{Average number of bugs for single mode.}{single_crash}
boringssl     &          0 &          0 &          0 & \textbf{1}  &          0  &          0 \\
c-ares        &          1 & \textbf{2} & \textbf{2} &          1  & \textbf{2}  &          1 \\
guetzli       &          0 &          0 &          0 &          0  &          0  &          0 \\
lcms          &          0 &          0 &          0 &          0  &          0  &          0 \\
libarchive    &          0 &          0 &          0 &          0  &          0  &          0 \\
libssh        &          0 &          0 &          0 & \textbf{1}  &          0  &          0 \\
libxml2       &          0 & \textbf{1} &          0 & \textbf{1}  & \textbf{1}  &          0 \\
openssl-1.0.1 &          0 &          0 &          0 &          0  &          0  &          0 \\
openssl-1.0.2 & \textbf{2} &          1 &          0 &          1  &          1  & \textbf{2} \\
openssl-1.1.0 &          0 &          0 &          0 &          0  &          0  &          0 \\
pcre2         & \textbf{2} &          1 &          1 &          1  & \textbf{2}  &          1 \\
proj4         &          0 &          0 &          0 & \textbf{1}  &          0  &          0 \\
re2           &          0 &          0 &          0 & \textbf{1}  &          0  &          0 \\
woff2         &          0 &          0 &          0 & \textbf{1}  &          0  &          0 \\
freetype2     &          0 &          0 &          0 &          0  &          0  &          0 \\
harfbuzz      &          0 &          0 &          0 & \textbf{1}  &          0  &          0 \\
json          & \textbf{1} & \textbf{1} &          0 &          0  & \textbf{1}  &          0 \\
libjpeg       &          0 &          0 &          0 &          0  &          0  &          0 \\
libpng        &          0 & \textbf{1} & \textbf{1} & \textbf{1}  & \textbf{1}  & \textbf{1} \\
llvm          &          0 &          0 & \textbf{1} & \textbf{1}  &          0  & \textbf{1} \\
openthread    &          0 &          0 &          0 & \textbf{1}  &          0  &          0 \\
sqlite        &          0 &          0 &          0 & \textbf{1}  & \textbf{1}  & \textbf{1} \\
vorbis        &          1 &          1 & \textbf{2} &          1  &          1  & \textbf{2} \\
wpantund      &          0 &          0 &          0 &          0  &          0  &          0 \\  
\midrule  
Total         & 7 & 8 & 7 & \textbf{15} & 10 & 9 \\  
\end{mytable_single}

\vspace{0.2cm}
%It is reasonable to draw the conclusion that
\noindent\textbf{In conclusion: }Different base fuzzers perform variously on distinct target applications, showing the diversity for the base fuzzers. The more diversity of these base fuzzers, the more differently they perform on different applications. Furthermore, the above three types of effects should be considered and could be incorporated into the fuzzing evaluation guideline \cite{klees2018evaluating} to avoid biased test cases or metrics selection when evaluating different types of fuzzing optimization.

\section{Does performance vary in different modes? }
    
\begin{figure*}[!htb]
\centering
\begin{minipage}[!htbp]{0.19\textwidth}
     \centering
     \includegraphics[width=1.0\textwidth]{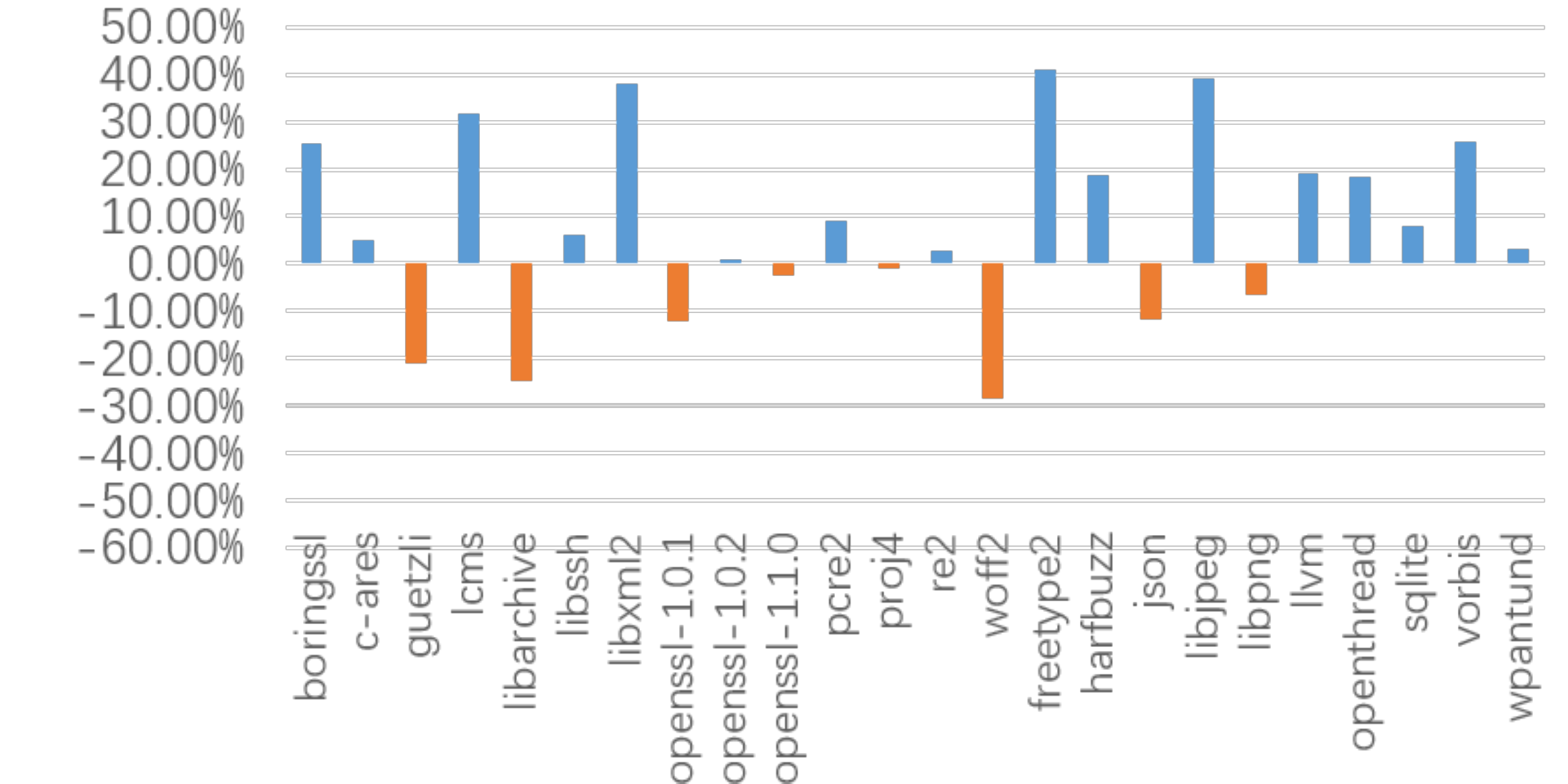}
     \small{(a) performance of AFLFast in single thread}
\end{minipage}
\begin{minipage}[!htbp]{0.19\textwidth}
     \centering
     \includegraphics[width=1.0\textwidth]{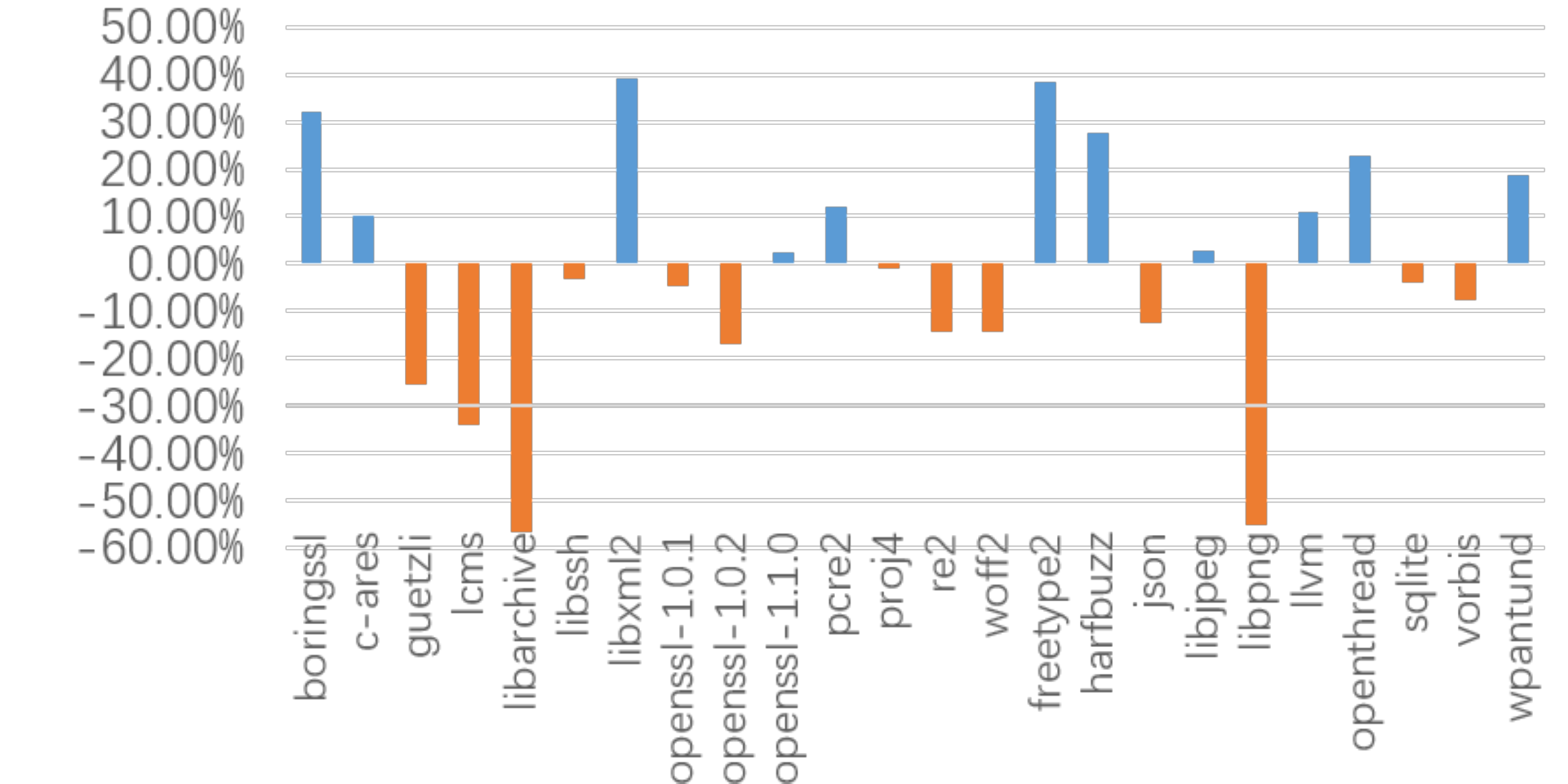}     
     \small{(b) performance of FairFuzz in single thread}
\end{minipage}
\begin{minipage}[!htbp]{0.19\textwidth}
     \centering
     \includegraphics[width=1.0\textwidth]{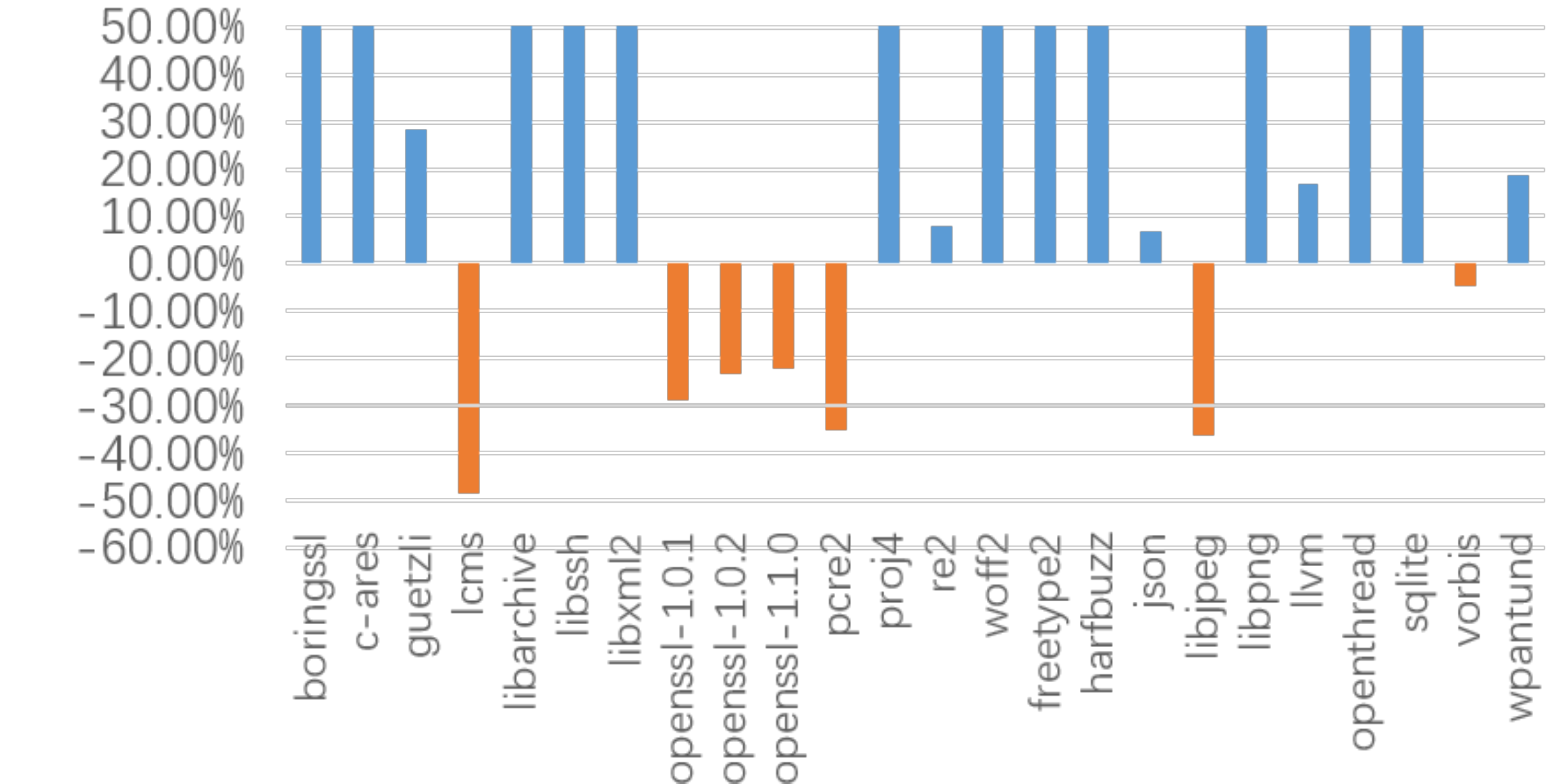}     
     \small{(c) performance of libFuzzer  in single thread}
\end{minipage}
\begin{minipage}[!htbp]{0.19\textwidth}
     \centering
     \includegraphics[width=1.0\textwidth]{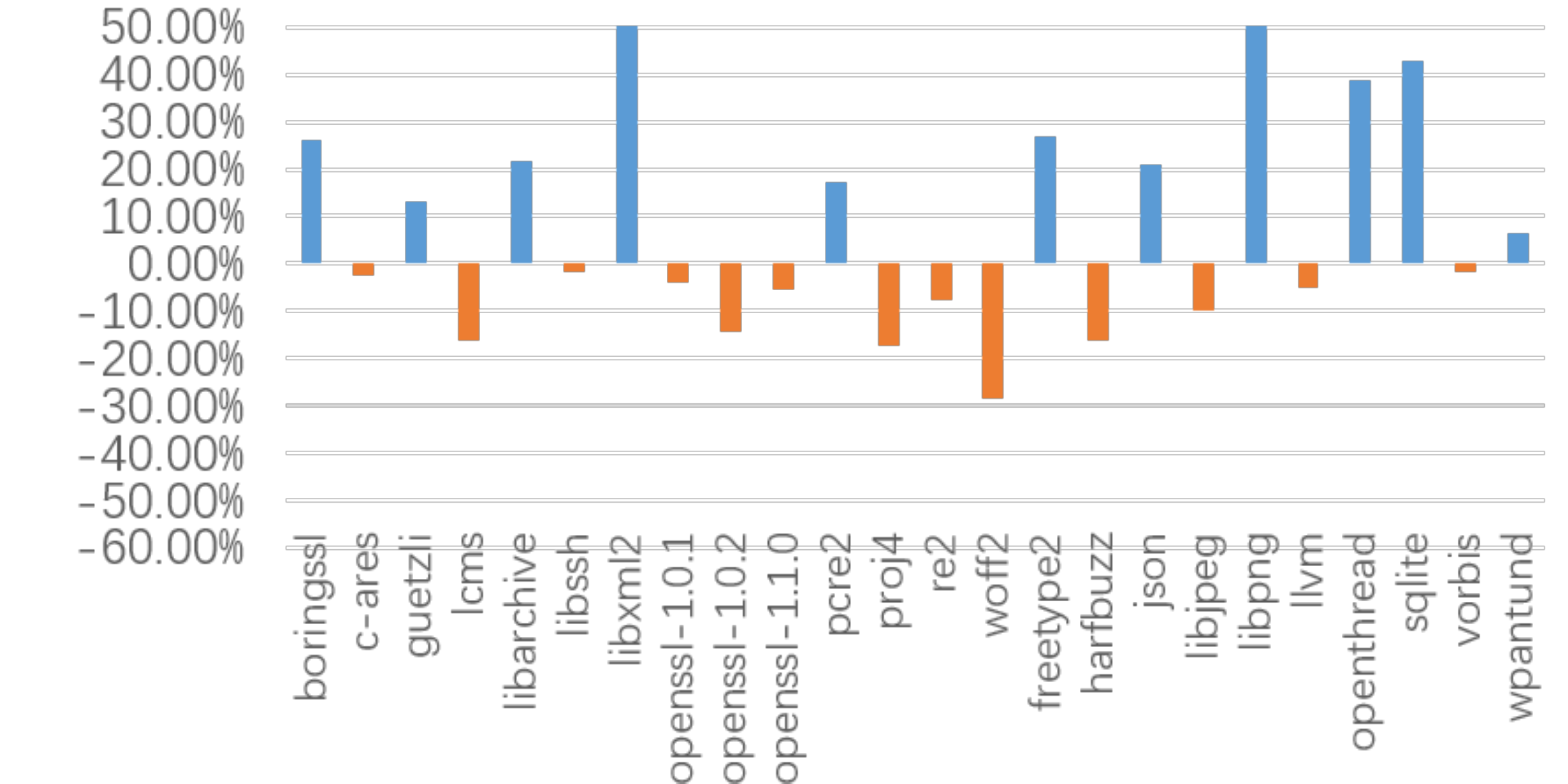}     
     \small{(d) performance of Radamsa  in single thread}
\end{minipage}
\begin{minipage}[!htbp]{0.19\textwidth}
     \centering
     \includegraphics[width=1.0\textwidth]{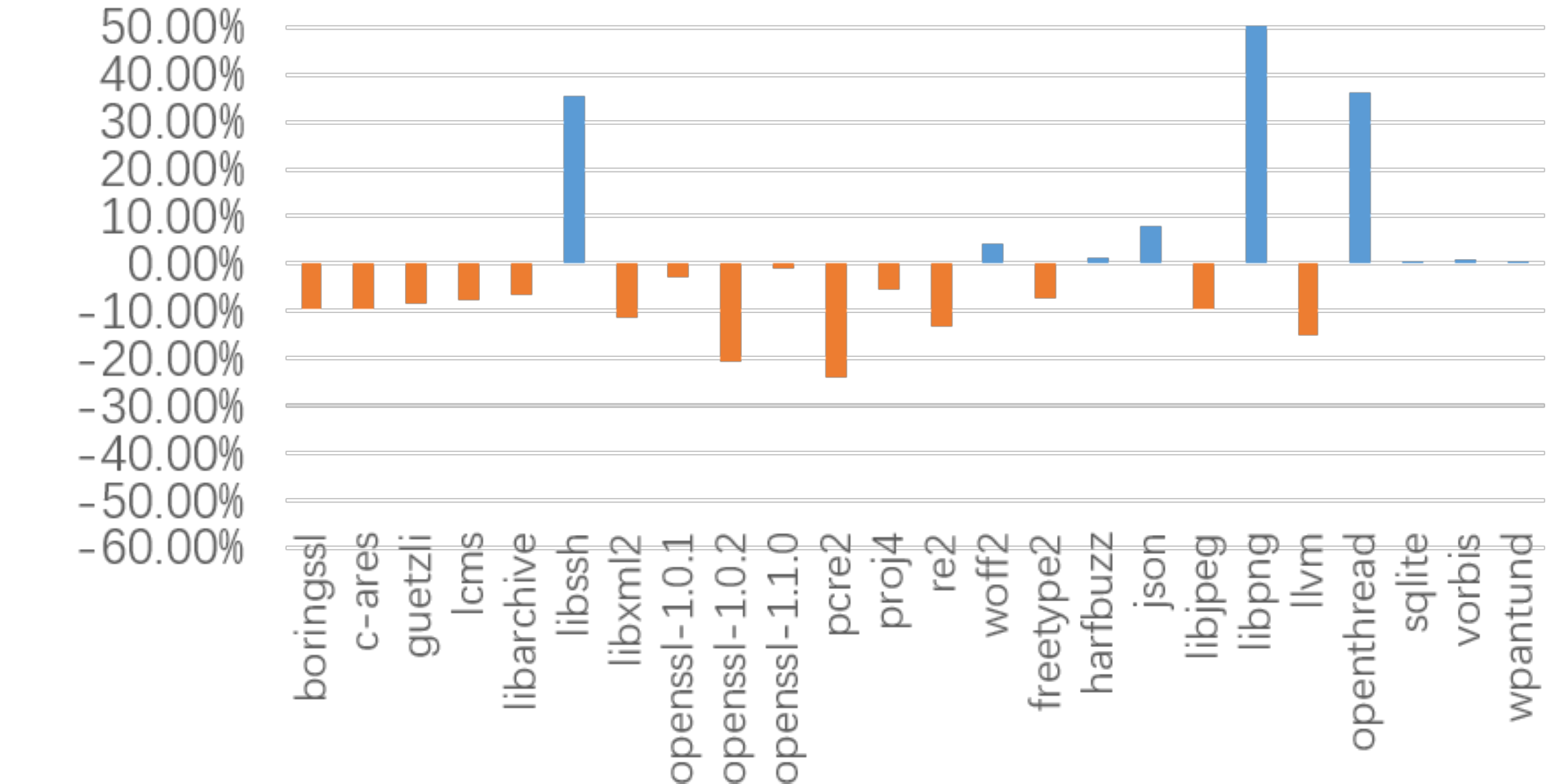}     
     \small{(d) performance of QSYM  in single thread}
\end{minipage}
\caption{Paths covered by base fuzzers compared with AFL in single mode on a single core.} %, blue means improvement and yellow means worse
\label{fig:paths executed compared with AFL}
\end{figure*}

%\begin{figure*}[!htbp]
%\centering
%\begin{minipage}[!htbp]{0.24\textwidth}
%     \centering
%     \includegraphics[width=1.0\textwidth]{img/branch_AFLFast.pdf}
%     \small{(a) performance of AFLFast in single thread}
%\end{minipage}
%\begin{minipage}[!htbp]{0.24\textwidth}
%     \centering
%     \includegraphics[width=1.0\textwidth]{img/branch_FairFuzz.pdf}     
%     \small{(b) performance of FairFuzz in single thread}
%\end{minipage}
%\begin{minipage}[!htbp]{0.24\textwidth}
%     \centering
%     \includegraphics[width=1.0\textwidth]{img/branch_LibFuzzer.pdf}     
%     \small{(c) performance of libFuzzer  in single thread}
%\end{minipage}
%\begin{minipage}[!htbp]{0.24\textwidth}
%     \centering
%     \includegraphics[width=1.0\textwidth]{img/branch_Radamsa.pdf}     
%     \small{(d) performance of Radamsa  in single thread}
%\end{minipage}
%\caption{Branches covered by base fuzzers compared with AFL in single mode on a single core, blue means improvement and yellow means worse.}
%\label{fig:paths executed compared with AFL}
%\end{figure*}

\begin{figure*}[!htb]
\centering
\begin{minipage}[!htbp]{0.19\textwidth}
     \centering
     \includegraphics[width=1.0\textwidth]{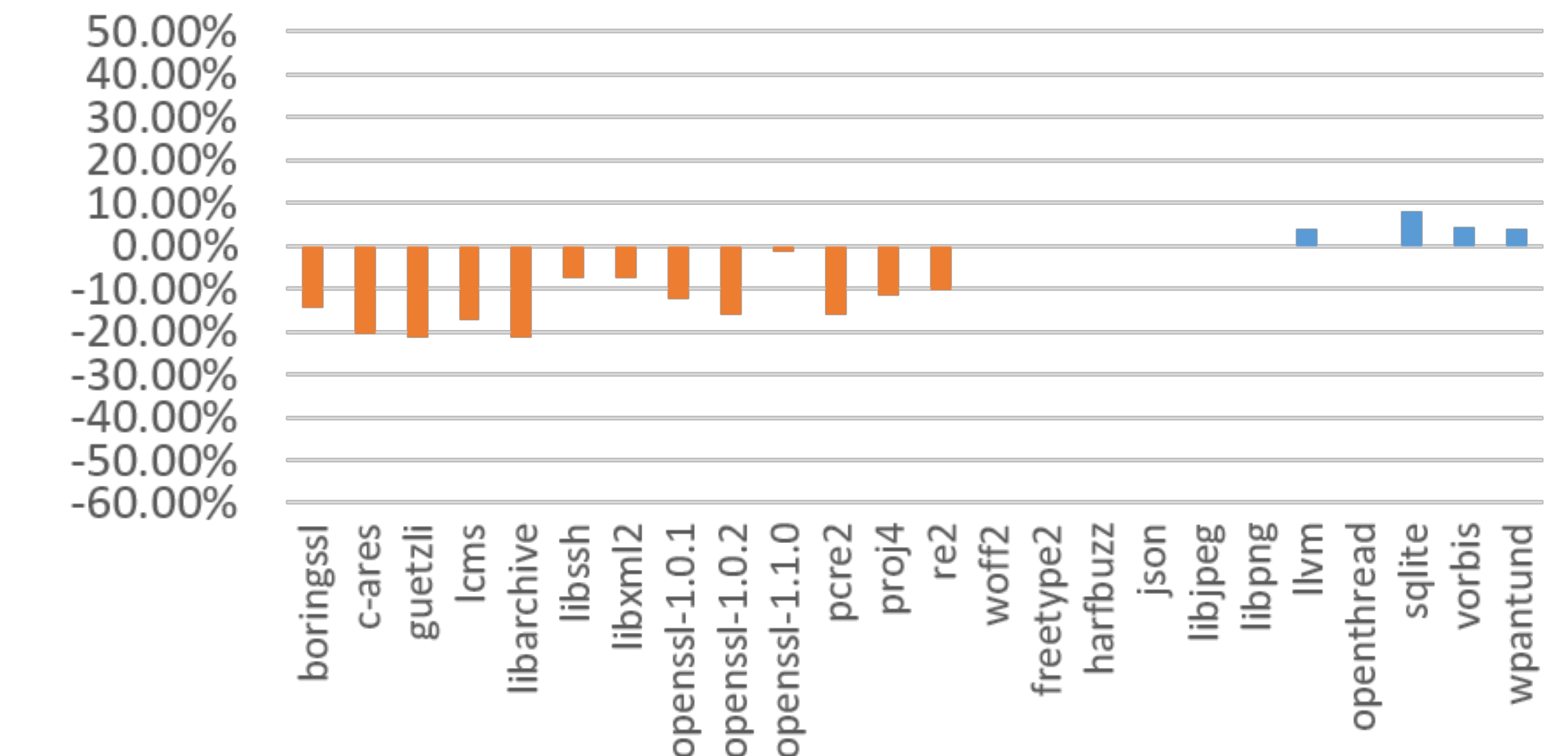}
     \small{(a) performance of AFLFast in four threads}
\end{minipage}
\begin{minipage}[!htbp]{0.19\textwidth}
     \centering
     \includegraphics[width=1.0\textwidth]{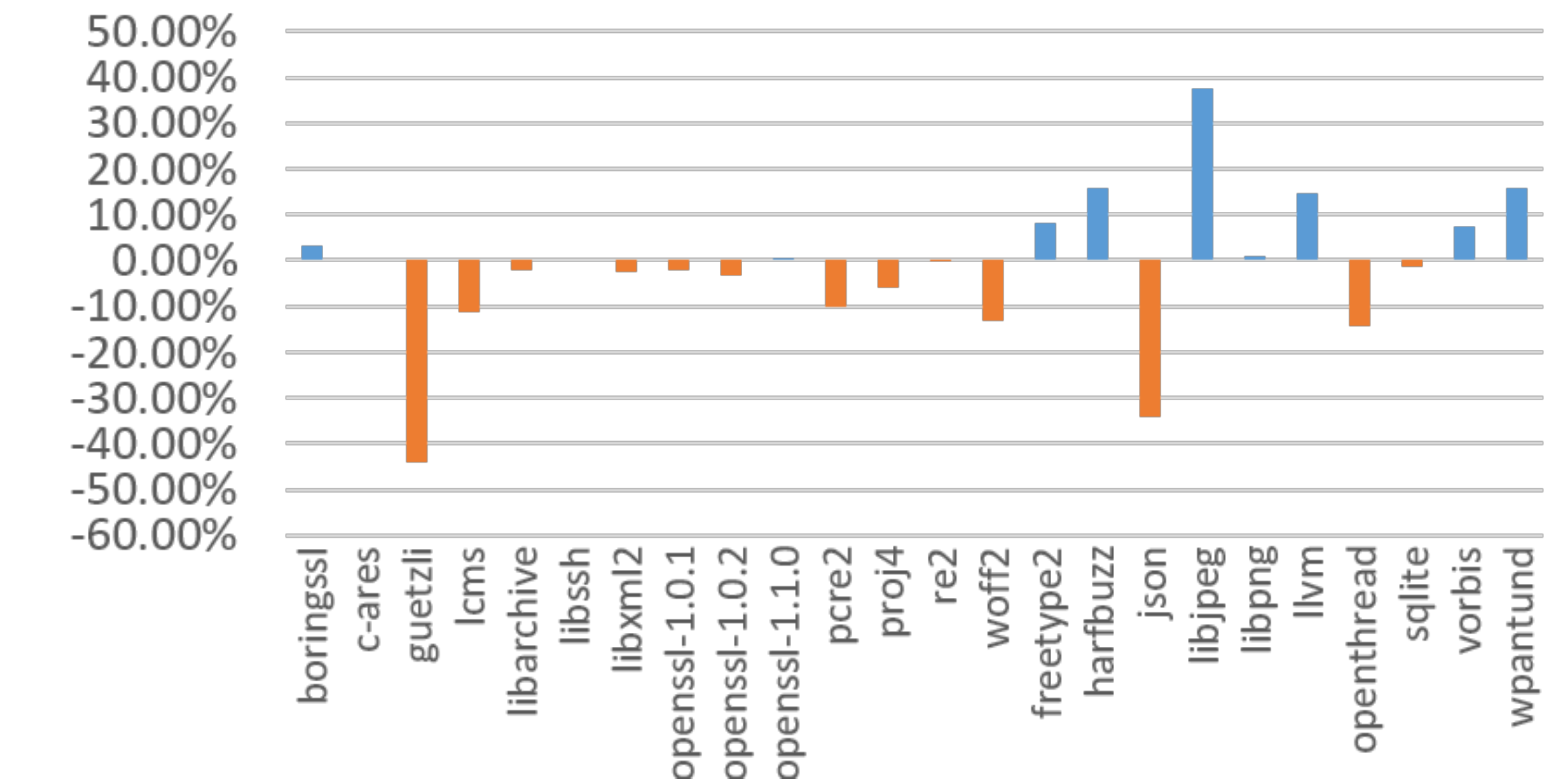}     
     \small{(b) performance of FairFuzz in four threads}
\end{minipage}
\begin{minipage}[!htbp]{0.19\textwidth}
     \centering
     \includegraphics[width=1.0\textwidth]{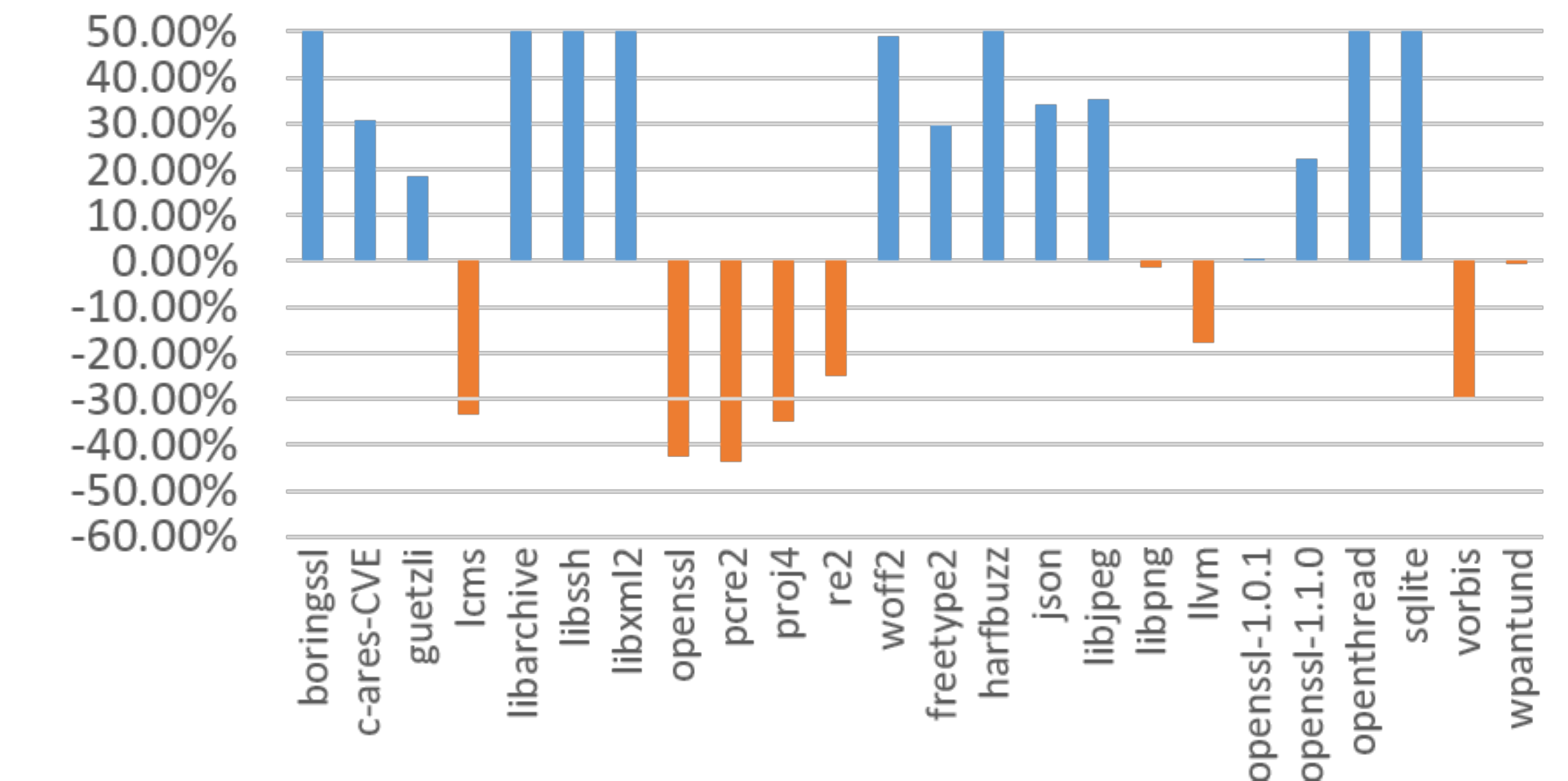}     
     \small{(c) performance of libFuzzer in four threads}
\end{minipage}
\begin{minipage}[!htbp]{0.19\textwidth}
     \centering
     \includegraphics[width=1.0\textwidth]{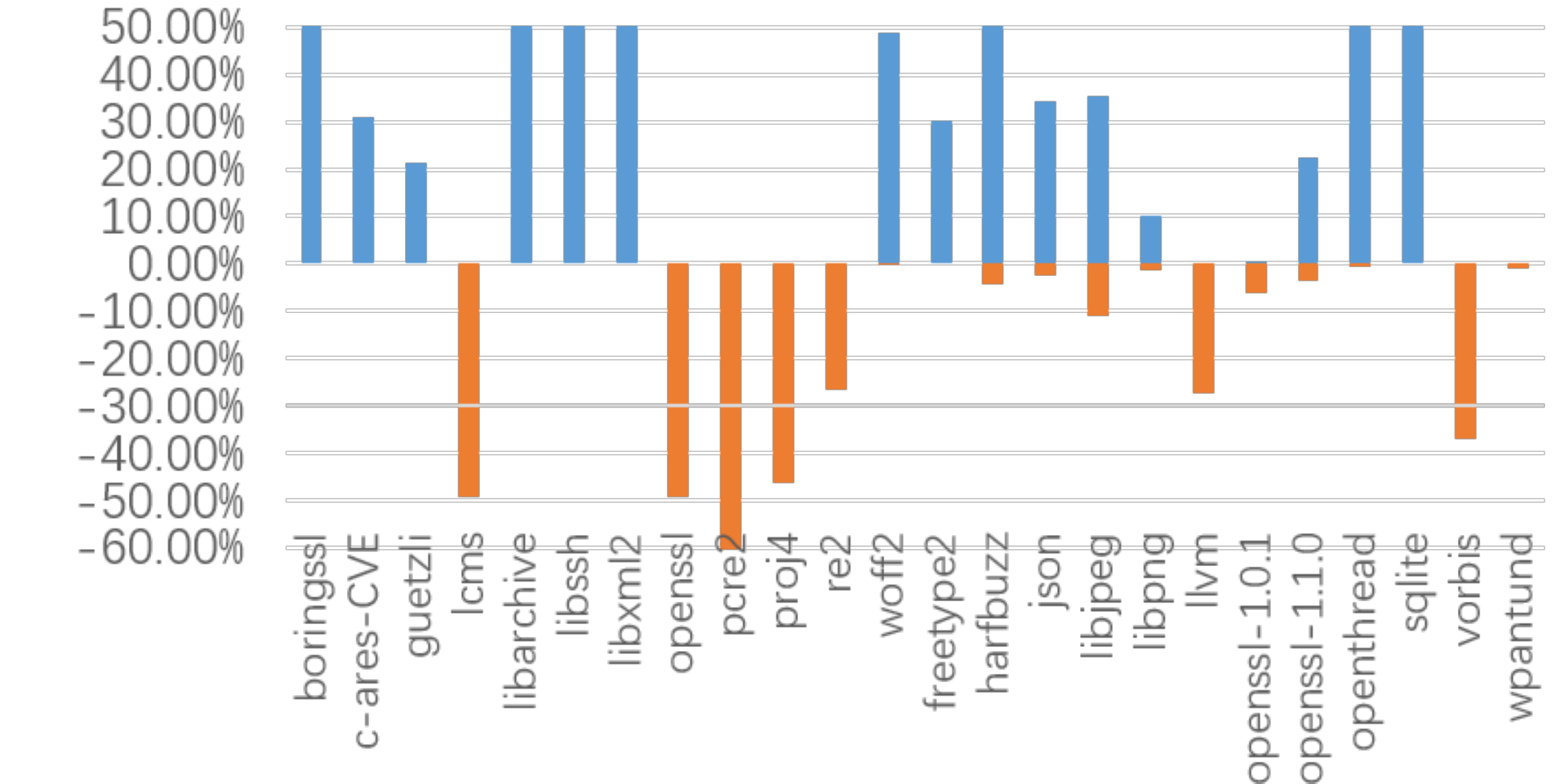}     
     \small{(d) performance of Radamsa in four threads}
\end{minipage}
\begin{minipage}[!htbp]{0.19\textwidth}
     \centering
     \includegraphics[width=1.0\textwidth]{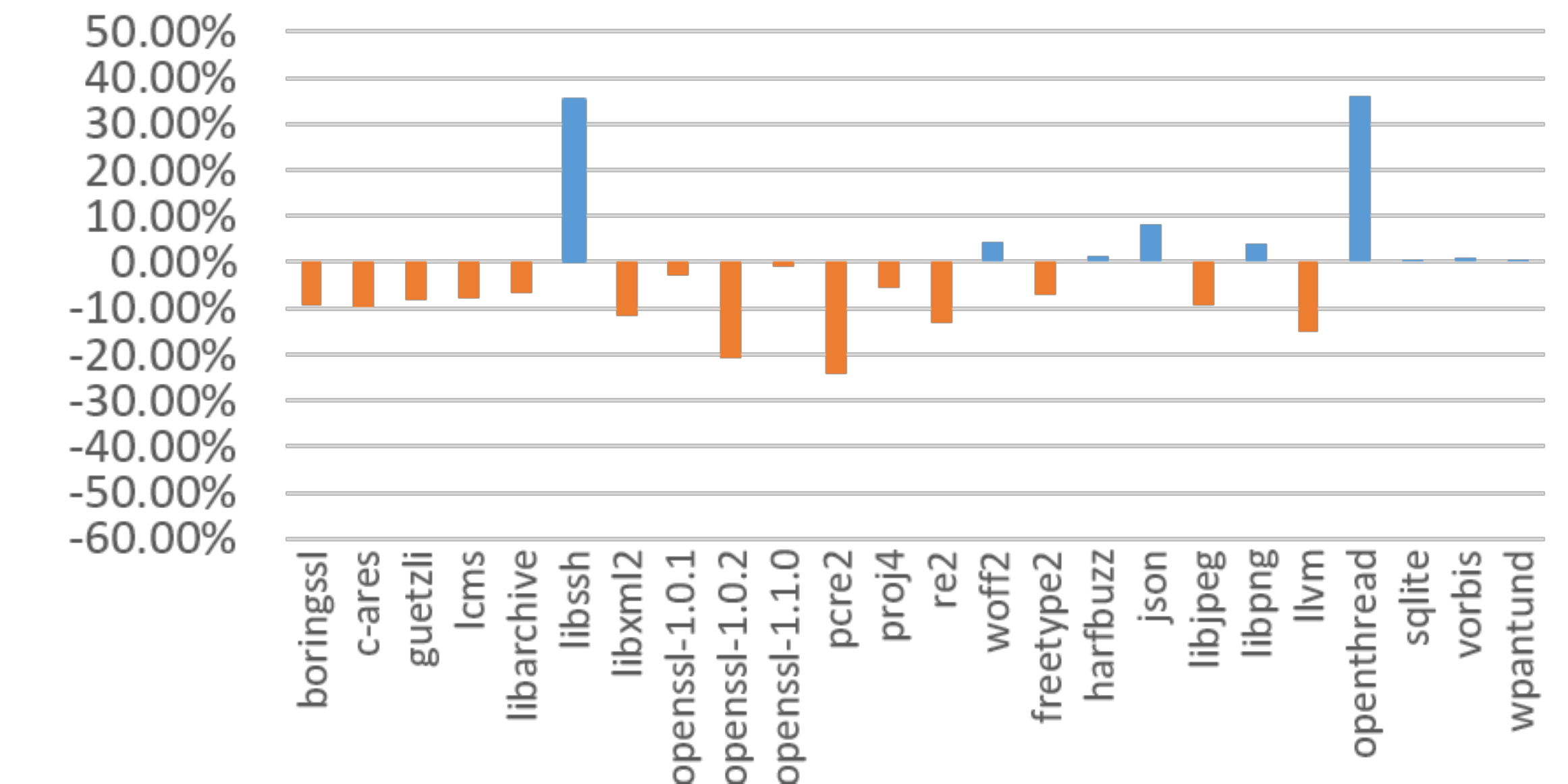}     
     \small{(d) performance of QSYM in four threads}
\end{minipage}
\caption{Paths covered by base fuzzers compared with AFL in parallel mode with four threads on four cores.}
\label{fig:paths-AFL4}
\end{figure*}

\begin{figure*}[!htb]
\centering
\begin{minipage}[!htbp]{0.19\textwidth}
     \centering
     \includegraphics[width=1.0\textwidth]{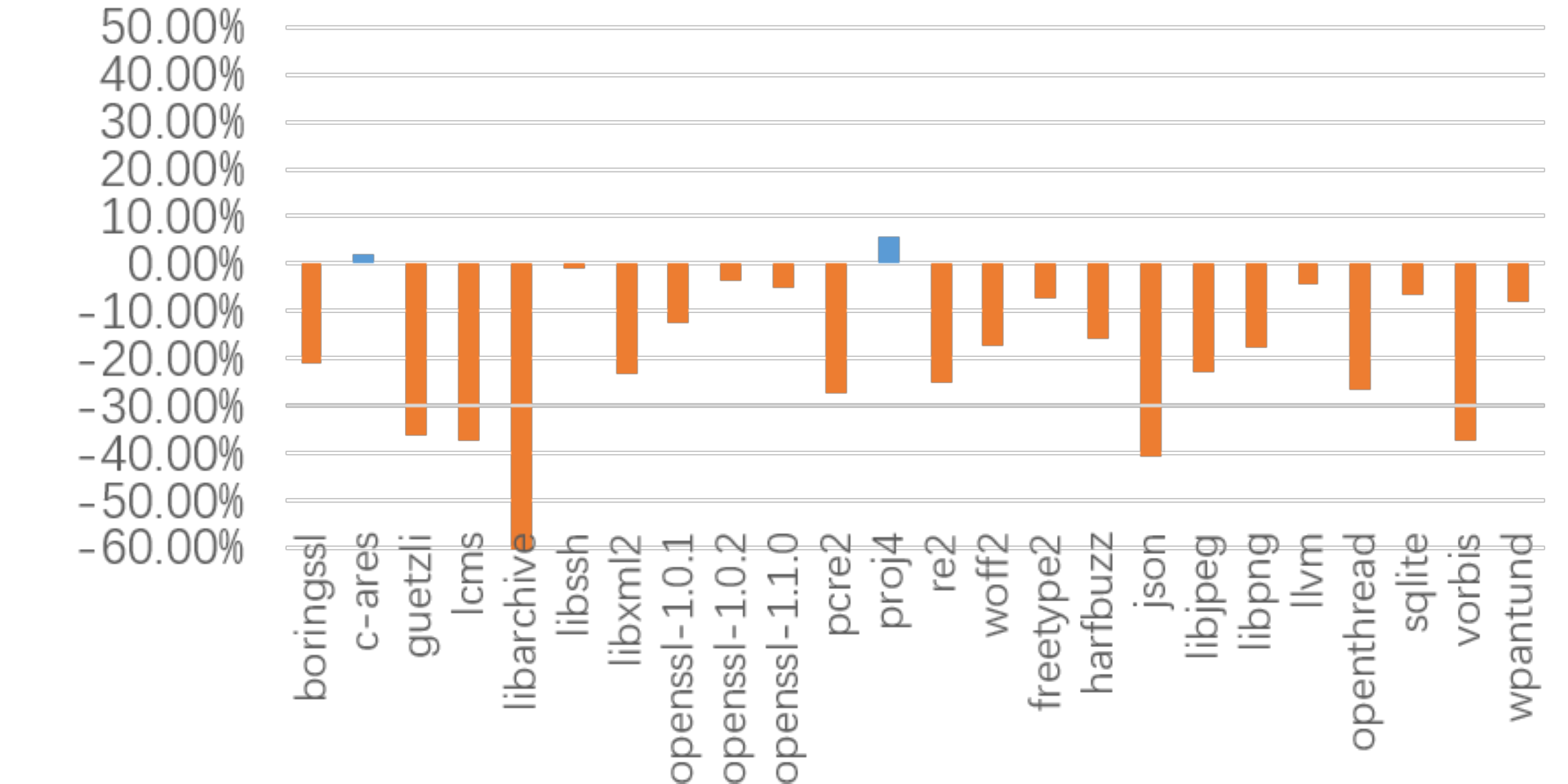}     
     \small{(a)  performance of \toolFour ~in four threads}
\end{minipage}
\begin{minipage}[!htbp]{0.19\textwidth}
     \centering
     \includegraphics[width=1.0\textwidth]{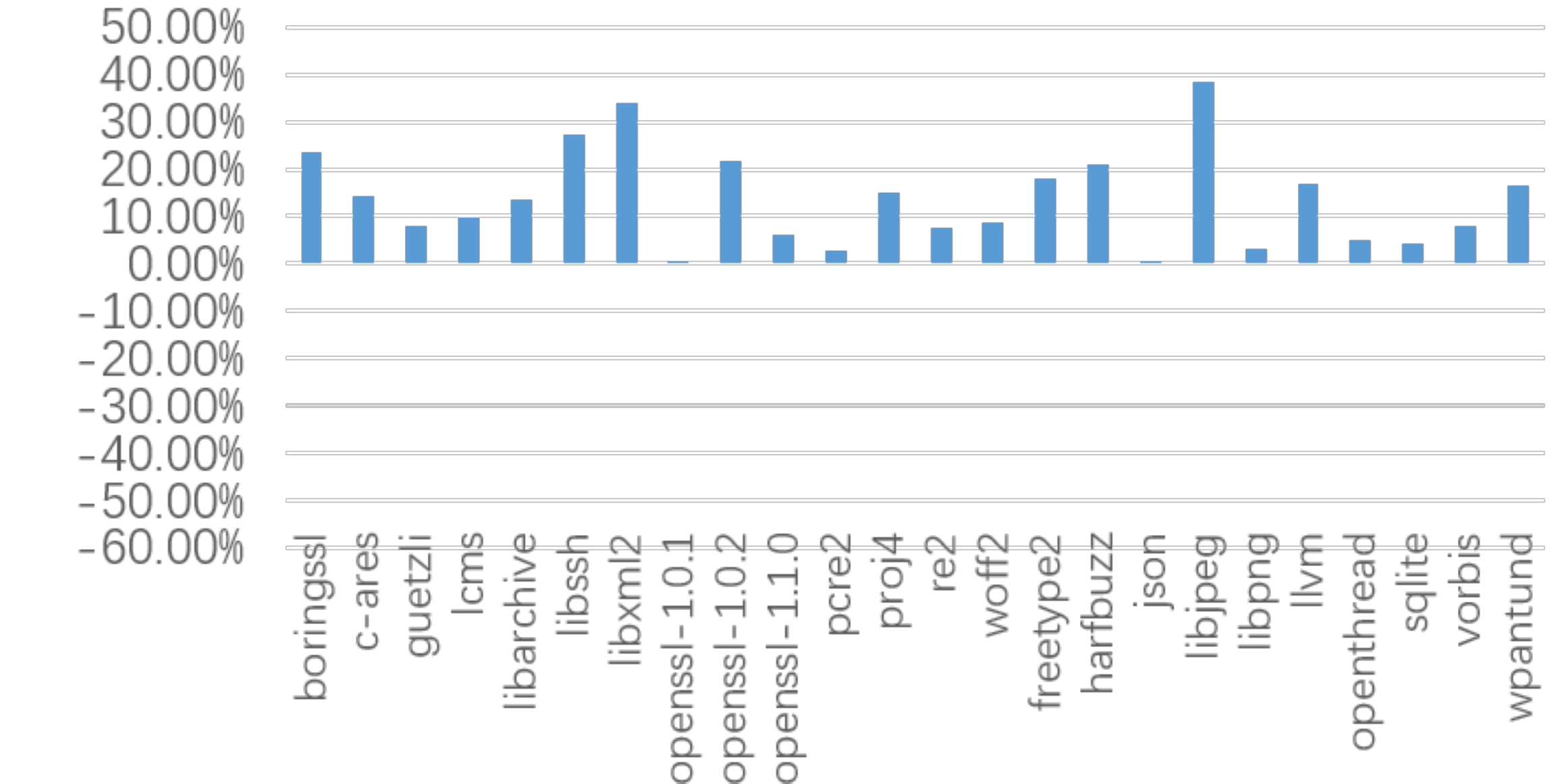}     
     \small{(a)  performance of \toolOne ~in four threads}
\end{minipage}
\begin{minipage}[!htbp]{0.19\textwidth}
     \centering
     \includegraphics[width=1.0\textwidth]{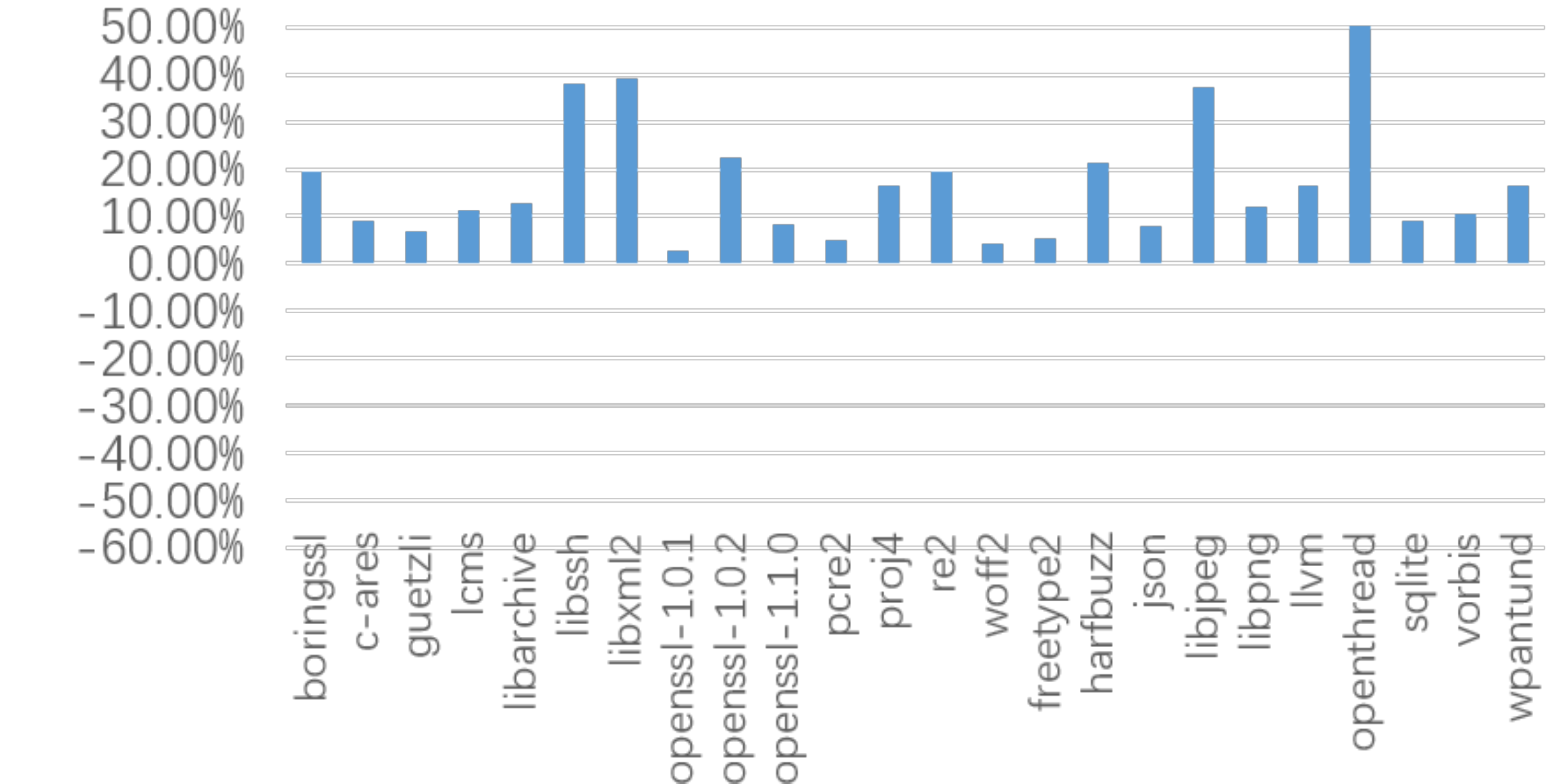}     
     \small{(a)  performance of \toolFive ~in four threads}
\end{minipage}
\begin{minipage}[!htbp]{0.19\textwidth}
     \centering
     \includegraphics[width=1.0\textwidth]{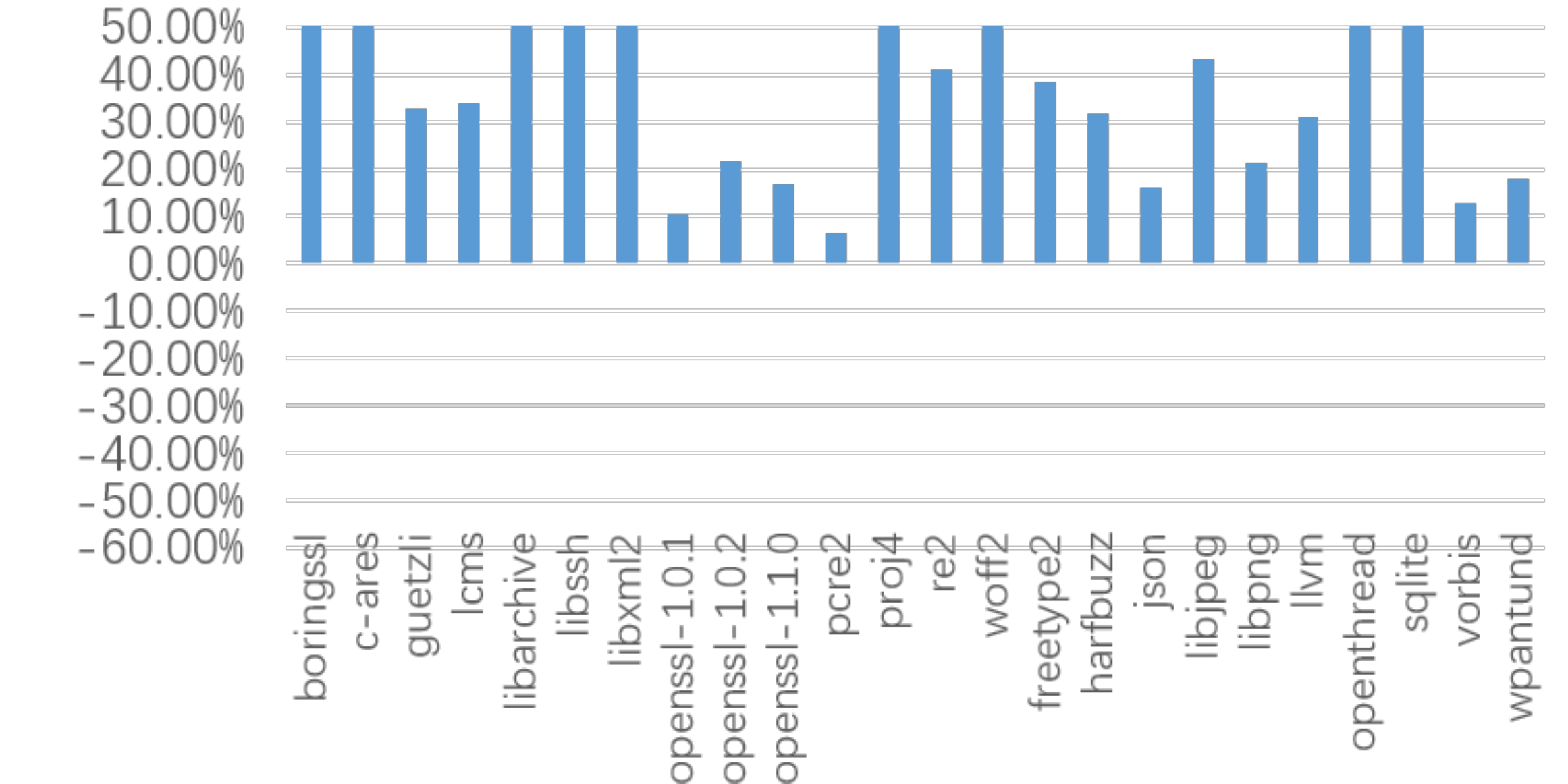}     
     \small{(b) performance of \toolTwo ~in four threads}
\end{minipage}
\begin{minipage}[!htbp]{0.19\textwidth}
     \centering
     \includegraphics[width=1.0\textwidth]{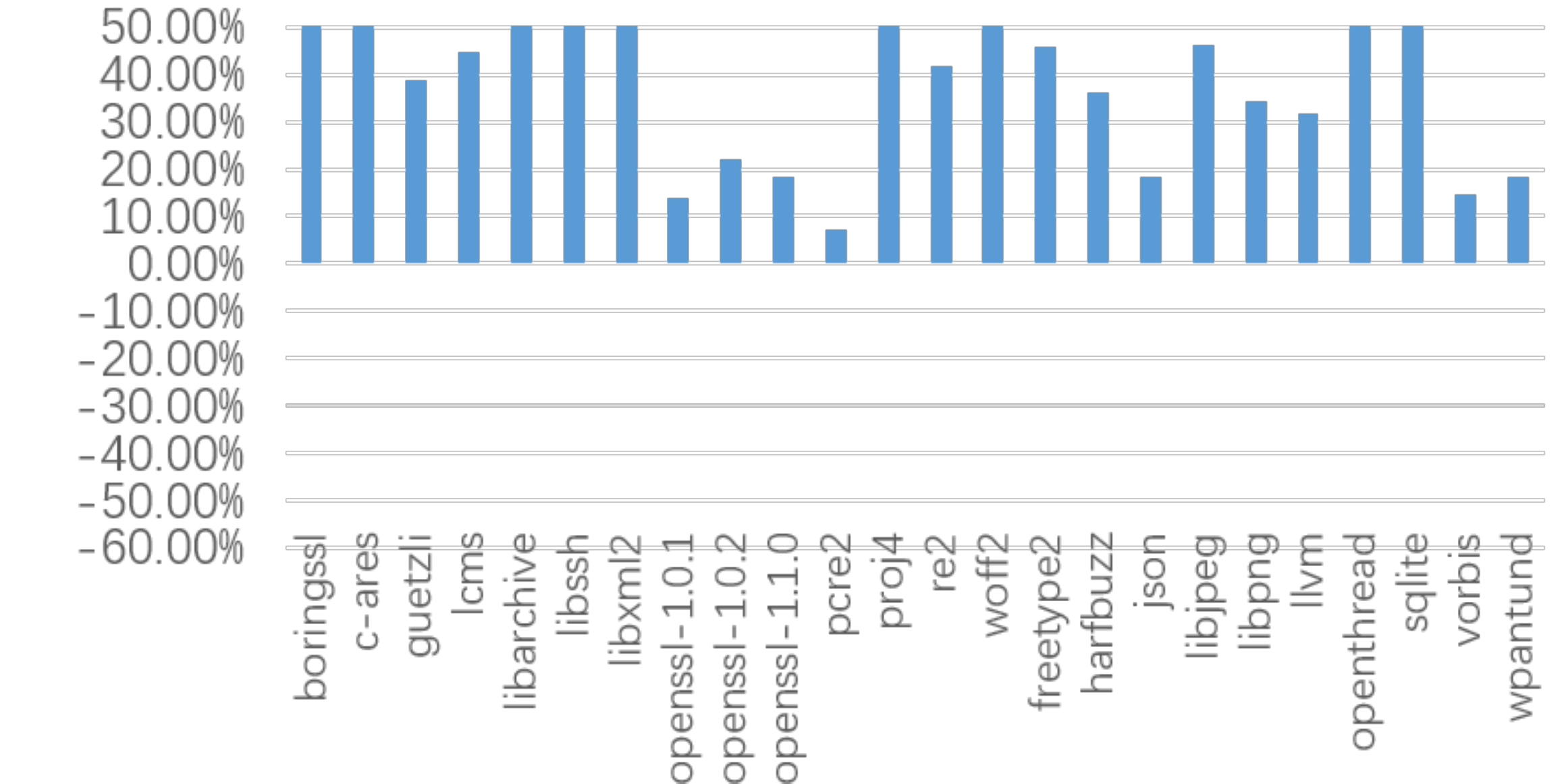}     
     \small{(c) performance of \toolThree ~in four threads}
\end{minipage}
\caption{Paths covered by \EnFuzz ~with four threads on four cores compared with AFL in parallel mode with four threads on four cores. \toolFour ~without the proposed seed synchronization performs the worst, and \EnFuzz ~performs the best.}
\label{fig:paths executed by EnFuzz}
\end{figure*}

%\begin{figure*}[!htbp]
%\centering
%\begin{minipage}[!htbp]{0.3\textwidth}
%     \centering
%     \includegraphics[width=1.0\textwidth]{img/path_enfuzz1.pdf}     
%     \small{(a)  performance of Enfuzz1 in four threads}
%\end{minipage}
%\begin{minipage}[!htbp]{0.3\textwidth}
%     \centering
%     \includegraphics[width=1.0\textwidth]{img/path_enfuzz2.pdf}     
%     \small{(b) performance of Enfuzz2 in four threads}
%\end{minipage}
%\begin{minipage}[!htbp]{0.3\textwidth}
%     \centering
%     \includegraphics[width=1.0\textwidth]{img/path_enfuzz3.pdf}     
%     \small{(b) performance of Enfuzz3 in four threads}
%\end{minipage}
%\caption{Branches covered by \EnFuzz ~with four threads on four cores compared with AFL in parallel mode with four threads on four cores.}
%\label{fig:paths executed by EnFuzz}
%\end{figure*}

%During this evaluation, we get many interesting results that are missed or not consistent with many previous literature studies. 
We choose AFL as the baseline, and compare other tools with AFL on path coverage to demonstrate the performance variation. 
Figure \ref{fig:paths executed compared with AFL} shows the average number of paths executed on Google's fuzzer-test-suite by each base fuzzer compared with AFL in single mode.
%Considering that \toolFour, \toolOne, \toolTwo ~and \toolThree ~use four times of computing resources compared with any base fuzzer running in single mode, for fairness,
We also collect the result of each base fuzzer running in parallel mode with four threads, 
and the result is presented in Figure \ref{fig:paths-AFL4}.
Figure \ref{fig:paths executed by EnFuzz} shows the average number of paths executed by \EnFuzz ~compared with AFL in parallel mode with four CPU cores.
From these results, we get the following conclusions:
\begin{itemize}

%1. 不同base fuzzer 在不同项目上的fuzzing效果差异性很大，不稳定 （在单线程和多线程场景下都有这问题）， 说明它们的generalization ability 不好
\item From the results of Figure \ref{fig:paths executed compared with AFL} and Figure \ref{fig:paths-AFL4}, we find that compared with AFL, the two optimized fuzzers AFLFast and FairFuzz, block coverage guided fuzzer libFuzzer, generation-based fuzzer Radamsa and hybrid fuzzer QSYM perform variously on different applications both in single mode and in parallel mode. It demonstrates that the performance of these base fuzzers is challenged by the diversity of the diverse real applications. %at least not as good as the descriptions in their corresponding literature, which shows that their performance is almost always better than that of AFL. 
The performance of their fuzzing strategies cannot constantly perform better than AFL. The performance variation exists in these state-of-the-art fuzzers. %, while in previous literature studies, fuzzers such as AFLFast and FairFuzz are evaluated to be always better than AFL within 24 hours performance. 

\item Comparing the result of Figure \ref{fig:paths executed compared with AFL} and Figure \ref{fig:paths-AFL4}, we find that the performance of these base fuzzers in parallel mode are quite different from those in single mode, especially for AFLFast and FairFuzz. In single mode, the other two optimized base fuzzers perform better than AFL in many applications. But in parallel mode, the result is completely opposite that the original AFL performs better on almost all applications. %, which is missed by evalutions of many previous literature studies.% The detail will be discussed later.

%2. 与base fuzzer相比，enfuzz 在所有的项目上都是最好的，说明通过ensemble的方法，能有效提升generalization ability
\item From the result of Figure \ref{fig:paths executed by EnFuzz}, it reveals that \toolOne, \toolTwo ~and \toolThree ~always perform better than AFL on the target applications.
For the same computing resources usage where AFL running in parallel mode with four CPU cores, \toolOne ~covers 11.26\% more paths than AFL, ranging from 4\% to 38\% in single cases, \toolFive ~covers 12.48\% more paths than AFL, ranging from 5\% to 177\% in single cases, \toolTwo ~covers 37.50\% more paths than AFL, ranging from 13\% to 455\% in single cases.
\toolThree ~covers 42.39\% more paths than AFL, ranging from 14\% to 462\% in single cases.
Through ensemble fuzzing, the performance variation can be reduced. % and the performance metrics can be improved. %, even with little diversity among base fuzzers.

%3. 通过多个enfuzz之间的比较，发现 集成的base fuzzers diversity越大,最终集成的效果越好
\item From the result of Figure \ref{fig:paths executed by EnFuzz}, it reveals that \toolFour ~without seed synchronization performs worse than AFL parallel mode under the same resource constraint. Compared with \toolOne, \toolFive ~covers 
1.09\% more paths, \toolTwo ~covers 23.58\% more paths. 
For \toolThree, it covers 27.97\% more paths than \toolOne, 26.59\% more paths than \toolFive,  3.6\% more paths than \toolTwo, and always performs the best on all applications. 
The more diversity among those integrated base fuzzers, the better performance of ensemble fuzzing, and the seed synchronization contributes more to the improvements.

\end{itemize}

\noindent\textbf{In conclusion: }the performance of the state-of-the-art fuzzers is greatly challenged by the diversity of those real-world applications, and it can be improved through the ensemble fuzzing approach. Furthermore, those optimized strategies work in single mode can not be directly scaled to parallel mode which is widely used in industrial practice. The ensemble fuzzing approach is a critical enhancement to the single and parallel mode of those optimized strategies.

\end{document}